\documentclass[preprint,12pt]{elsarticle}

\usepackage{amssymb}

\biboptions{sort&compress}
\begin{document}

\begin{frontmatter}



\title{Two-Dimensional Ordering of Bacteriorhodopsins in a Lipid Bilayer and Effects Caused by Repulsive Core between Lipid Molecules on Lateral Depletion Interaction: A Study based on a Thermodynamic Perturbation Theory}


\author[inst1]{Keiju Suda}

\affiliation[inst1]{organization={Department of Chemistry, Graduate School of Science, Kyushu University},
            addressline={Motooka 744, Nishi-ku}, 
            city={Fukuoka-shi},
            postcode={819-0395}, 
            state={Fukuoka},
            country={Japan}}

\author[inst2]{Ayumi Suematsu}
\author[inst1]{Ryo Akiyama}

\affiliation[inst2]{organization={Faculty of Science and Engineering, Kyushu Sangyo University},
            addressline={Matsukadai 2-31, Higashi-ku}, 
            city={Fukuoka-shi},
            postcode={813-8503}, 
            state={Fukuoka},
            country={Japan}}

\begin{abstract}
Using binary hard disk mixture models, we studied the two-dimensional ordering of bacteriorhodopsins in a lipid bilayer. The phase diagrams were calculated using the thermodynamic perturbation theory. We examined two types of effective interactions to discuss the lateral depletion effects caused by repulsive core interaction between lipid molecules. The results indicate that the core repulsions drastically broaden the coexistence region for the fluid--ordered phase.
\end{abstract}




\end{frontmatter}


\section{INTRODUCTION}
Bacteriorhodopsin (bR) is a transmembrane protein \cite{RHenderson,RHenderson2}. The wild-type bRs form trimers, and the trimers construct a two-dimensional ordering structure in the membrane of {\it Halobacterium salinarum}, which is an archaebacterium. The hexagonal ordering structure is not infinite, but it has been called a two-dimensional crystal of bR. There is no chemical bond between bRs \cite{review}, and the driving force of this ordering is unclear. \par
Our previous study\cite{Suda} suggested that the lateral depletion effect\cite{AO,AOpolymer,FMT1,FMT2,FMT3,NakamuraAkiyama,vrij,Lekkerkerker,Lekkerkerker2,Frenkel,Dijkstra1,Dijkstra2,Dijkstra3,Dijkstra4,Dijkstra5,Dijkstra6,Dijkstra7,Cinacchi,Lopez,Velasco,Suematsu2016,Chang} is dominant in the driving force of ordering. The study focused on the difference between the ``crystallization'' behaviors of wild-type bR and mutant bR, which construct the ordering structure. Although the mutant monomers do not construct trimers, they form a crystal\cite{monomer}. The critical concentrations (CCs) of the mutant bRs for the ``crystallization'' are much larger than those of the wild-type bRs. According to experiments by M. P. Krebs et al., the ratio ${\rm CCR(= CC1/CC3)}$ is about 10.2, where CC1 and CC3 are the critical concentrations for the mutant monomers and the wild-type trimers, respectively\cite{CCR}.\par
In the previous study \cite{Suda}, we calculated the phase diagram using simple theories for two-dimensional binary hard-disk systems. There is a reason for adopting this simple approach. Several researchers have adopted similar approaches in the study of three-dimensional binary hard-sphere systems and reported adequate results\cite{Lekkerkerker,Lekkerkerker2,Frenkel,Dijkstra1,Dijkstra2,Dijkstra3,Dijkstra4,Dijkstra5,Dijkstra6,Dijkstra7,Cinacchi,Lopez,Velasco,Suematsu2016}. Therefore, we applied the ideas to two-dimensional systems. There are some studies on the binary hard-disk system \cite{Rovert,disappear,Ott}. However, the size ratio was far from that for the lipid molecule/bR. In our studies, the size ratio was limited from lipid molecule/bR monomer to lipid molecule/bR trimer\cite{Suda}. The binary hard-disk system, composed of bR and lipid molecules, was connected with a lipid reservoir. We obtained the free energies for the fluid and ordered phases based on the free volume idea\cite{Lekkerkerker,Lekkerkerker2} and calculated the phase diagram. \par
The calculated results indicated that the depletion effect was too weak when the lipid molecules were modeled as an ideal gas, and it differed from the experimental result\cite{Suda}. On the other hand, we also described the pressure of the reservoir using the two-dimensional scaled particle theory (2D--SPT)\cite{SPT,2DSPT}, taking into account the effects of the repulsive core of the lipid molecule\cite{Suda}. We calculated the phase diagrams based on the models. The calculated phase diagrams showed that the repulsive core enhanced the depletion effect, and the calculated CCRs virtually agreed with the experimental results \cite{Suda}. Therefore, we discussed that the depletion effect significantly contributed to the ``crystallization''.\par
The reduced effective second virial coefficients for bRs were also calculated to discuss the depletion effects of the phase diagram calculation in a previous study\cite{Suda}. We examined two depletion interactions: namely, the Asakura--Oosawa (AO) \cite{AO,AOpolymer} and the modified Asakura--Oosawa (modified AO) interactions\cite{Suda}. The core repulsion between lipid molecules is ignored in the AO interaction. On the other hand, taking account of the core repulsion between the lipid molecules, we adopted 2D--SPT in the modified AO theory. As a result, the reduced effective second virial coefficients only for the modified AO interactions showed large negative values. Therefore, the phase diagrams and the reduced effective second virial coefficients were consistent. However, the phase behaviors of bR of this effective interaction have not been obtained in the previous paper\cite{Suda}.\par
In the previous simple approach\cite{Suda}, we obtained the phase diagrams without calculating the effective interaction between bRs. We study similar problems of CCR in the present paper, but the phase diagrams are now calculated using a completely different approach. First, we calculated the effective interactions between bR. Next, we calculated the free energies using the thermodynamic perturbation theory\cite{Suematsu2016,Velasco,TP} with the effective interactions and obtained the phase diagrams. Finally, the phase diagrams gave us CCRs.

\section{Model and Theory}
\ \ \ We adopted a binary hard-disk model in the present study. The model is the same as that in reference\cite{Suda}. The lipid bilayer was regarded as a condensed 2D plane space. The lipid molecules, bR monomers, and bR trimers were modeled as small, medium, and large hard disks, respectively. The diameter for the bR trimer $\sigma_{\rm tri}$ was estimated as 6.2 nm\cite{Suda,6.2}. The diameter for the bR monomer $\sigma_{\rm mono}$ was estimated as 3.0 nm\cite{Suda}. The diameter for the lipid molecule $\sigma_{\rm lip}$was estimated as 0.5 nm\cite{Suda,cell}.  In addition, the two lipid diameters $\sigma_{\rm lip}=$ 0.4 and 0.6 nm were also examined to remove the arbitrariness for the model. \par
The system consisted of lipid and bR molecules. The binary hard-disk system was in osmotic equilibrium with a reservoir. Therefore, we obtained the effective interactions between bR molecules and calculated the phase diagrams using the effective one-component system. The effective interactions between bR molecules are explained as follows. \par
The AO and modified AO potentials\cite{Suda} were adopted as the effective potential for bRs. The lipid molecules cannot enter the excluded area $V_{\rm ex}$ around the bR molecules in this model. The AO potential $\omega_{\rm AO}$ between bRs can be written as 
\begin{eqnarray}
\omega_{\rm AO}(\it r) &=&\infty,\ \ \it r<\sigma_{\rm bR},\\
\omega_{\rm AO}(\it r) &=& -\rho_{\rm lip}^{\rm res} \rm k_{B}\it{T}\rm{\Delta} \it{V}_{\rm ex}(\it r),\ \  \it r>\sigma_{\rm bR},
\end{eqnarray}
where $\rho_{\rm lip}^{\rm res}$ is the number density of lipid molecules in the reservoir, $\rm k_B$ is Boltzmann constant, \it T \rm{is} the absolute temperature, $\Delta V_{\rm ex}$ is the overlap area of excluded volumes, \it r \rm is the distance between centers of bRs, and $\sigma_{\rm bR}$ is the diameter for bR. $\Delta V_{\rm ex}$ is expressed analytically as follows, 
\begin{eqnarray}
\Delta V_{\rm ex} (r)=\frac{1}{2} \sigma_{\rm bR}^2 (1+q)^2 \rm{arccos} \left[\frac{\it r}{\sigma_{\rm bR} (1+\it q)}\right]\nonumber\\
-\frac{1}{4} \sigma_{\rm bR}^2 (1+q)^2 \rm{sin}\left[2\rm{arccos}\left[\frac{\it r}{\sigma_{\rm bR} (1+\it q)}\right]\right]  
\end{eqnarray}
where \it q \rm is the diameter ratio between the lipid molecule and bR ($\sigma_{\rm lip}/\sigma_{\rm bR}$). $\rho_{\rm lip}^{\rm res} \rm{k_B} \it T$
is regarded as the pressure of ideal gas in the reservoir. Therefore, the AO interaction is the pressure-area work for the two-dimensional ideal gas, and the area is $\Delta V_{\rm ex}$\cite{Lekkerkerker}.\par
In the conventional AO theory, the repulsive interactions between depletants are ignored, and the lipid molecules overlap. In other words, the pressure in the reservoir is that of the ideal gas. On the other hand, we replaced the pressure of the ideal gas reservoir with that estimated using the two-dimensional scaled particle theory (2D--SPT)\cite{Suda,Lekkerkerker}, taking account of the repulsive interactions between depletants. Thus, the pressure in the reservoir was expressed as follows:
\begin{equation}
p^{\rm res}=\frac{\rho_{\rm lip}^{\rm res}}{\left(1-\eta_{\rm lip}^{\rm res} \right)^2}  \rm{k_B}\it{T},
\label{pres}
\end{equation}
where $\eta_{\rm lip}^{\rm res}$ is the reservoir's packing fraction of lipid molecules. Here, we call this the modified AO model. The modified AO potential $\omega_{\rm Mod}$, therefore, was written as

\begin{eqnarray}
\omega_{\rm Mod}(\it r) &=&\infty,\ \ \it r<\sigma_{\rm bR},\\
\omega_{\rm Mod}(\it r) &=& -\frac{\rho_{\rm lip}^{\rm res}}{\left(1-\eta_{\rm lip}^{\rm res} \right)^2} \rm{k_B}\it{T}\rm{\Delta} \it{V}_{\rm ex}(\it r),\ \  \it r>\sigma_{\rm bR}.
\end{eqnarray}
In this model, there are only two bRs in the lipid condensed system. That is, we assumed a very low bR packing fraction in the system. As the bR packing fraction increases, the lipid packing fraction in the system decreases. Therefore, the effective interaction depends on the bR packing fraction. Here, we adopted an approximation that $\omega_{\rm Mod}$ is independent of the bR packing fraction\cite{Suematsu2016,Velasco}.\par

\begin{figure*}[ht]
\centerline{
{\includegraphics[clip, width=8.9cm]{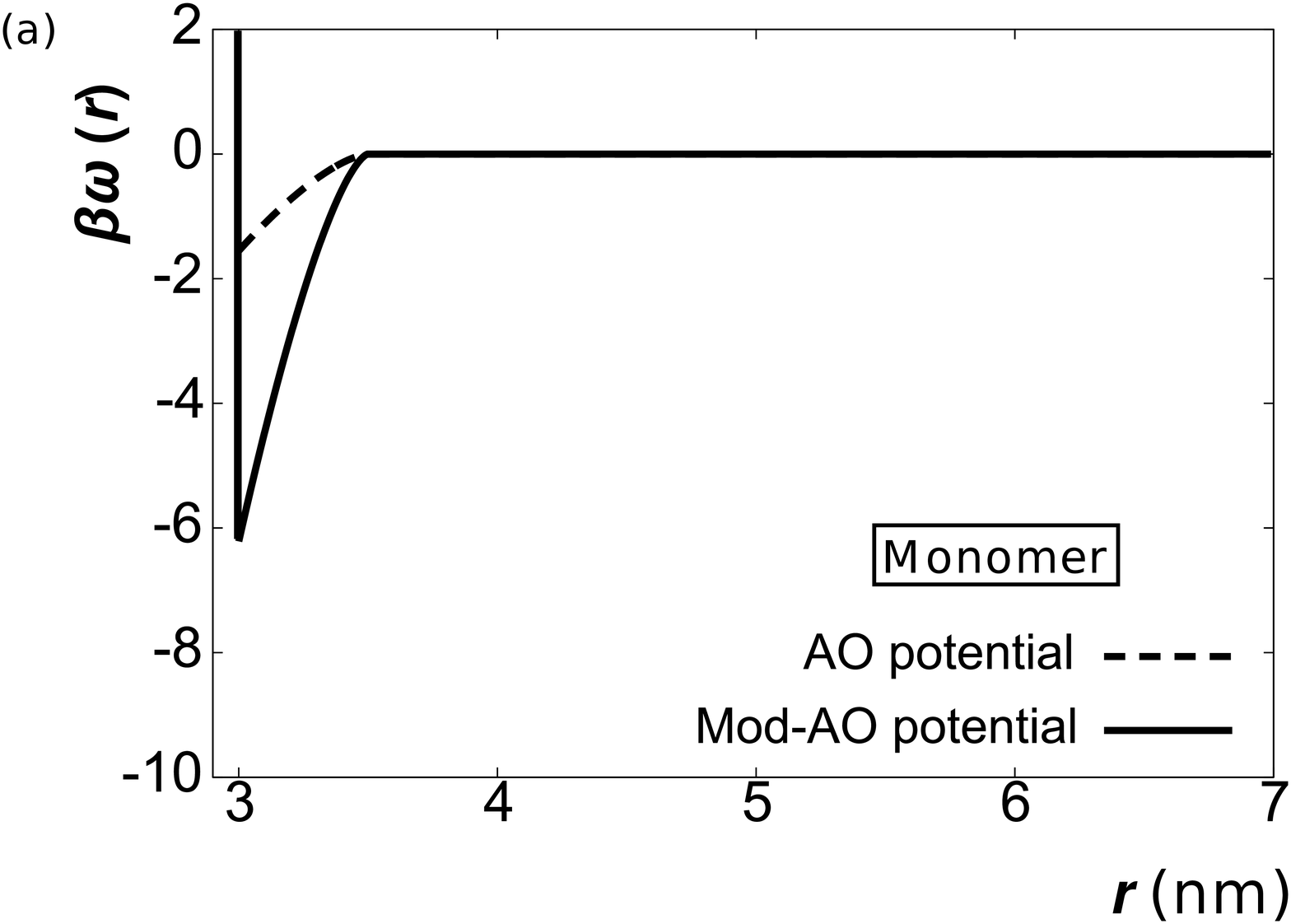}
\label{fig1A}}
\hfil
{\includegraphics[clip, width=8.9cm]{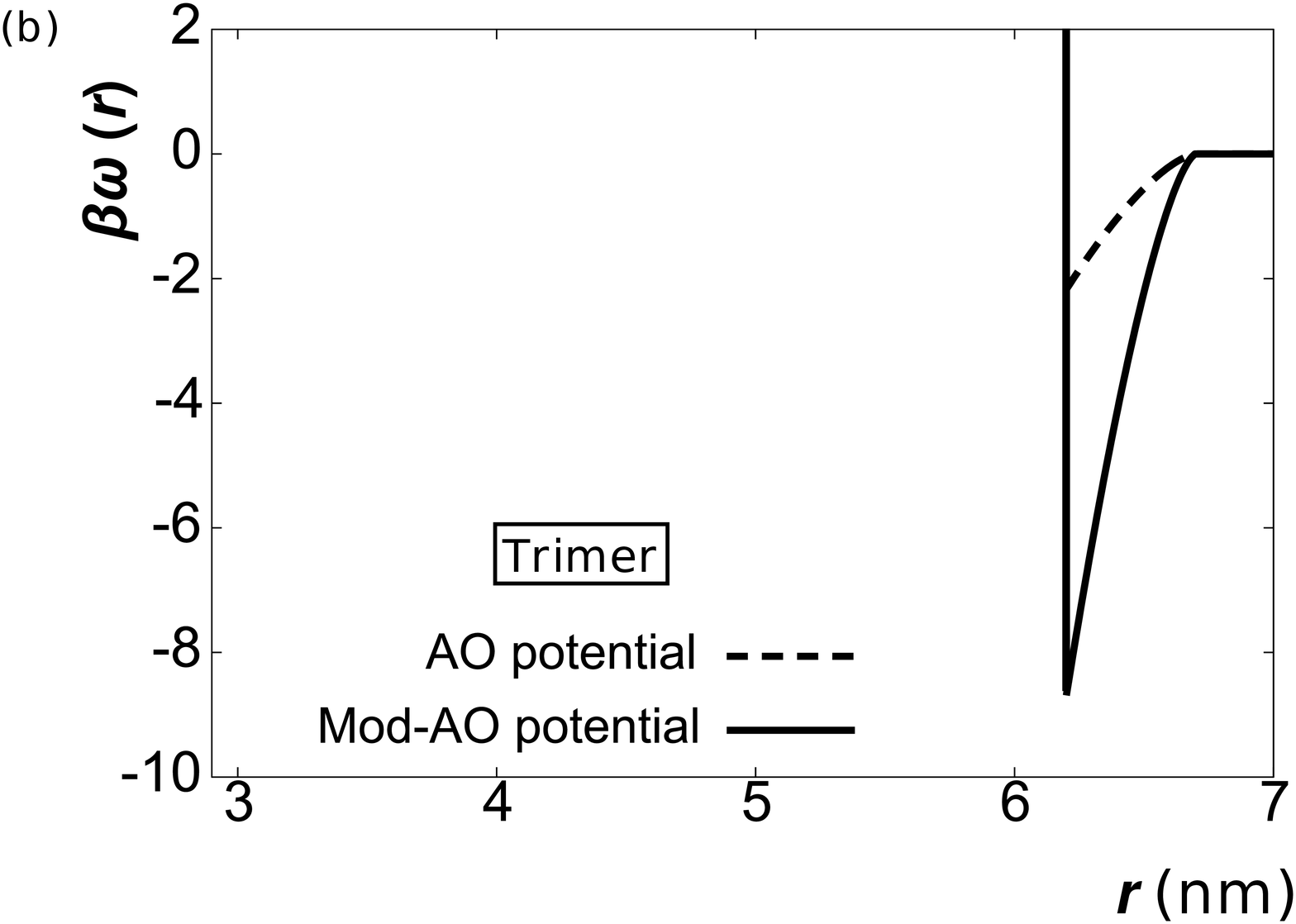}
\label{fig1B}}
}
\caption{The effective potential between bR monomers (a) and trimers (b) at $\eta_{\rm lip}^{\rm res} =0.5$. The lipid—bR diameter ratios, \it{q}\rm{s}, are 0.16667 and 0.08065, respectively. The dashed curves show the AO potential. The solid curves show the modified AO potential.}
\label{potential}
\end{figure*}

FIG. \ref{potential} shows the effective potentials at $\eta_{\rm lip}^{\rm res} =0.5$. The stability at the contact distance for the modified AO potential is much larger than that for the AO potential. The ratio is larger than three in the case of the bR monomers (FIG. \ref{potential}(a)). That is, the effective attraction for the modified AO potential is much stronger than that for the AO potential. In addition, the stability ratio becomes larger as the size ratio \it{q} \rm{decreases}. Thus, the stability ratio becomes larger than four in the case of the trimer (FIG. \ref{potential}(b)).\par
Here, we mention the difference between the AO and the modified AO potentials. We prepared the pressure in calculating the quasistatic PV-work to change the excluded volume of bRs for the lipid fluid. In the case of the AO potential, the pressure of the ideal gas was used. On the other hand, we adopted 2D--SPT in the case of the modified AO potential. The latter pressure is higher than the former because of the packing effect of the lipid hard disks. Therefore, the stability of the contact bR dimer for the modified AO potential is much larger than that for the AO potential.\par

To obtain the phase diagrams, we adopted the thermodynamic perturbation theory. In the perturbation theory, the Helmholtz free energy for the effective one-component system $F(N_{\rm bR}, V, T)$ was expressed as follows: 
\begin{eqnarray}
\frac{\beta F\left(N_{\rm bR},V,T\right)v_{\rm bR}}{V}=\frac{\beta F_0 \left(N_{\rm bR},V,T\right)v_{\rm bR}}{V}
+4\beta \eta_{\rm bR}^2 \int_0^\infty g_0 \left(\frac{r}{\sigma_{\rm bR}} \right)\omega\left(\frac{r}{\sigma_{\rm bR}} \right)  \frac{r}{\sigma_{\rm bR}}  d\left(\frac{r}{\sigma_{\rm bR}} \right),
\end{eqnarray}
where $V$ is the area of the system, $v_{\rm bR}$ is the area of one bR, $F_0$ is the Helmholtz free energy for the pure bR system, $\eta_{\rm bR}$ is the packing fraction of the bRs, $\omega(\frac{r}{\sigma_{\rm bR}} )$ is the effective potential between bRs, and $g_0 (\frac{r}{\sigma_{\rm bR}} )$ is the radial distribution function for the pure bR system scaled by $\sigma_{\rm bR}$.  \par
In the previous paragraph, the calculation of the perturbation part needed the radial distribution function. The radial distribution functions were obtained using the event chain Monte Carlo simulation\cite{hexatic}. The simulation box has $128^2$ hard disks, and the sampling was carried out over {$1.2\times 10^{11}$} steps after equilibration. The radial distribution functions were calculated within the ranging from $\eta_{\rm bR} =0.001$ to $\eta_{\rm bR} =0.905$ in the interval of $\eta_{\rm bR} =0.001$. The AO or the modified AO potential were substituted in $\omega(\frac{r}{\sigma_{\rm bR}} )$.\par
$F_0$ was obtained by thermodynamics as follows: 
\begin{equation}
\frac{\beta F_0 v_{\rm bR}}{V}=\beta \eta_{\rm bR} \mu_{\rm bR}^0-\beta p^0 v_{\rm bR},
\label{F0}
\end{equation}
where $\mu_{\rm bR}^0$ and $p^0$ are the chemical potential and the pressure for the pure bR system. The chemical potential and the pressure for the pure bR fluid phase, $\mu_{\rm f}^0$ and $p_{\rm f}^0$, were obtained by 2D-Carnahan-Staring like equation of state (2D--CSE)\cite{Suda,RHenderson}, as follows:

\begin{equation}
\beta \mu_{\rm f}^0=\rm{ln}\left[\frac{\Lambda^2}{\it{v}_{\rm bR} }\right]+\rm{ln}\left[\eta_{\rm bR}\right]-\frac{7}{8} \rm{ln}\left[1-\eta_{\rm bR} \right]+\frac{7}{8\left(1-\eta_{\rm bR} \right)} +\frac{9}{8\left(1-\eta_{\rm bR} \right)^2} -2,\label{muf}
\end{equation}

\begin{equation}
\beta p_{\rm f}^0 v_{\rm bR}=\frac{\eta_{\rm bR}+\frac{\left(\eta_{\rm bR} \right)^2}{8}}{\left(1-\eta_{\rm bR} \right)^2} ,
\end{equation}
where $\Lambda = \rm{h}(2\pi \it{m}_{\rm bR}\rm{k_B}\it{T})^{\rm -1/2}$ is the thermal de Broglie wavelength in the 2D space.  h and $m_{\rm bR}$ are the Planck constant and the mass for a bR, respectively. The chemical potential and the pressure for the pure bR ordered phase, $\mu_{\rm ord}^0$ and $p_{\rm ord}^0$, were obtained by cell theory\cite{Suda,2DFVT} as follows:
\begin{equation}
\beta\mu_{\rm ord}^0=\rm{ln}\left[\frac{\Lambda^2}{\it{v}_{\rm bR}}  \right]-2\rm{ln}\left[\frac{\eta_{\rm cp}}{\eta_{\rm bR}}-1 \right]+\frac{2\eta_{\rm cp}}{\eta_{\rm cp}-\eta_{\rm bR}}, \label{mus}
\end{equation}

\begin{equation}
\beta p_{\rm ord}^0 v_{\rm bR}=Z\frac{2\eta_{\rm bR}}{1-\frac{\eta_{\rm bR}}{\eta_{\rm cp}}},
\end{equation}
where $\eta_{\rm cp}=\pi/(2\sqrt{3})\approx 0.907$ is the packing fraction at close packing. The first term on the right-hand side, $\rm{ln}\left[\frac{\Lambda^2}{\it{v}_{\rm bR}} \right]$, in eq. (\ref{muf}), and eq. (\ref{mus}) does not affect the phase diagram, because the term is common in the fluid and ordered phases. We adopted the common tangent method to obtain the phase diagrams using free energy curves. The common tangent was drawn on the calculated free energy curves for the fluid and ordered phase. \par
In our previous study\cite{Suda}, the phase diagrams for the binary hard-disk system were obtained using two-dimensional free volume theory(2D--FVT). The semi-grand potential $\Omega(N_{\rm bR},V,T,\mu_{\rm lip})$ was calculated by 2D--FVT as follows:
\begin{equation}
\Omega(N_{\rm bR},V,T,\mu_{\rm lip})=F_0(N_{\rm bR},V,T)-p^{\rm res}\left< V_{\rm free}^{\rm mix} \right>_0,
\label{Omega}
\end{equation}
where $\left< V_{\rm free}^{\rm mix} \right>_0$ is the free volume for the lipid molecule in a pure bR system. $\left< V_{\rm free}^{\rm mix} \right>_0$ was calculated by 2D--SPT\cite{Suda,Lekkerkerker}. The equations for $F_0$ and $p^{\rm res}$ are eq. (\ref{F0}) and eq. (\ref{pres}). The pressure and chemical potential for bR were obtained from $\Omega(N_{\rm bR},V,T,\mu_{\rm lip})$. The phase diagrams were obtained using the following two equations:
\begin{eqnarray}
p_{\rm f}\left( \eta_{\rm bR}^{\rm fluid}, \eta_{\rm lip}^{\rm res}  \right) &=& p_{\rm ord}\left( \eta_{\rm bR}^{\rm ordered}, \eta_{\rm lip}^{\rm res}  \right),\label{p} \\
\mu_{\rm f,bR}\left( \eta_{\rm bR}^{\rm fluid}, \eta_{\rm lip}^{\rm res}  \right) &=& \mu _{\rm ord,bR}\left( \eta_{\rm bR}^{\rm ordered}, \eta_{\rm lip}^{\rm res}  \right),\label{mu}
\end{eqnarray}
where $p_{\rm f}$ and $p_{\rm ord}$ are the pressures for the fluid and ordered phases, $\eta_{\rm bR}^{\rm fluid}$ and $\eta_{\rm bR}^{\rm ordered}$ are the bR packing fractions for the fluid and ordered phases, $\eta_{\rm lip}^{\rm res}$ is the packing fraction of lipid for the reservoir, and $\mu_{\rm f,bR}$ and $\mu_{\rm ord,bR}$ are the bR chemical potentials for the fluid and ordered phases.

\section{Results}
\begin{figure*}[ht]
\centerline{
{\includegraphics[clip, width=8.9cm]{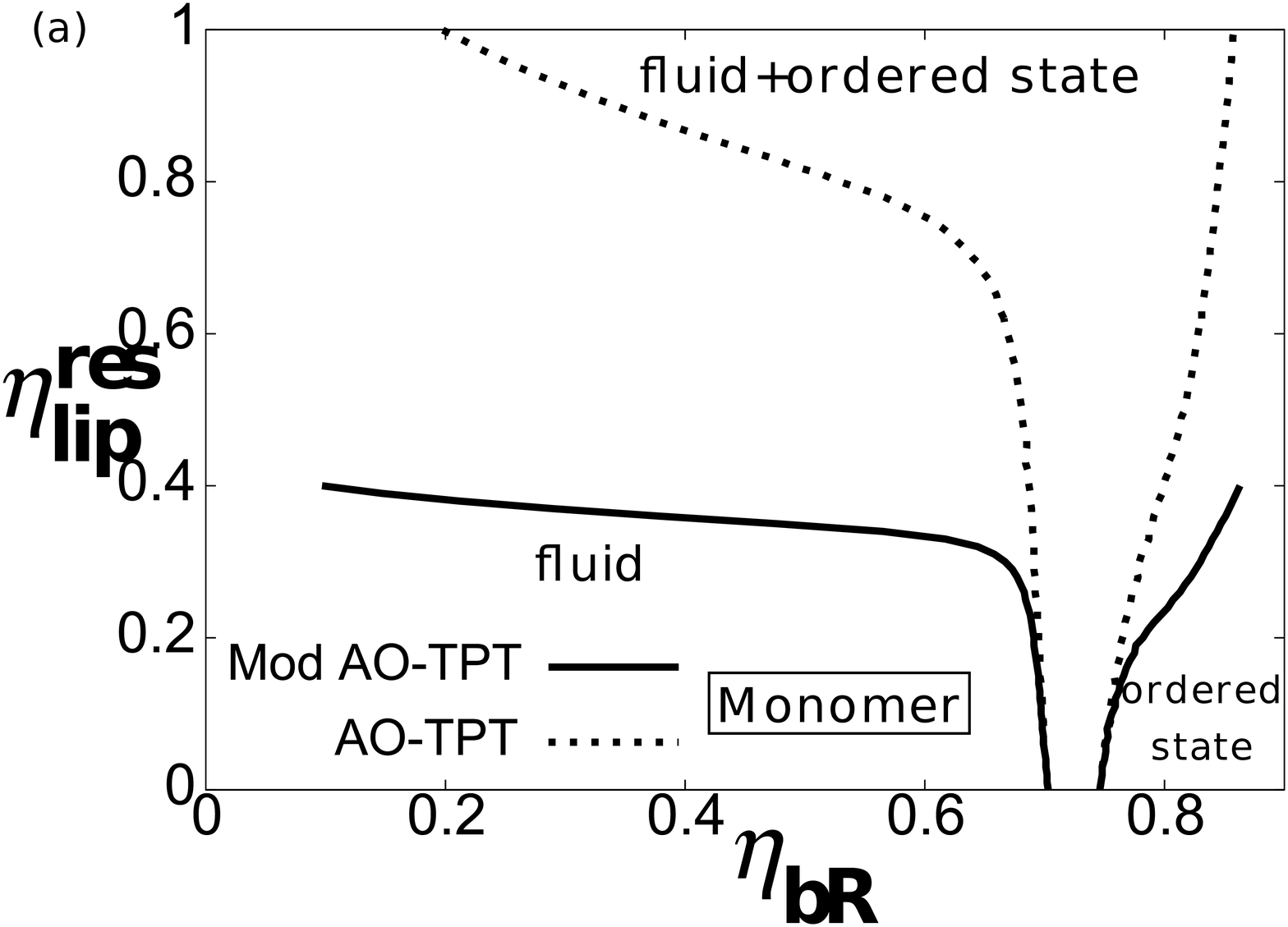}
\label{fig1A}}
\hfil
{\includegraphics[clip, width=8.9cm]{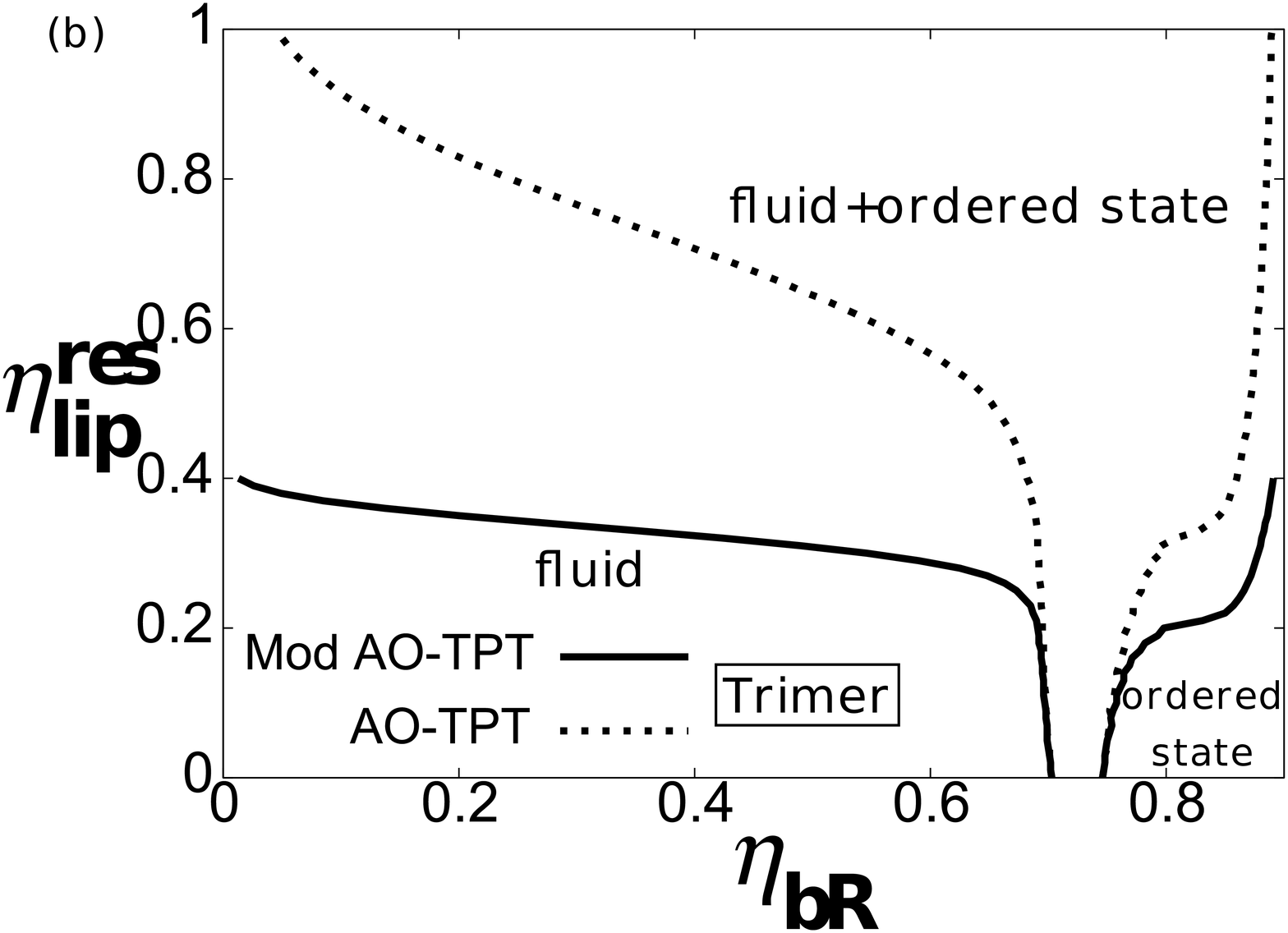}
\label{fig1B}}
}
\caption{Phase diagrams for (a) bR monomer ($q=0.16667$) and (b) trimer ($q=0.08065$). The dotted curves show the phase diagram calculated by the thermodynamic perturbation theory(TPT) with the AO potential. The solid curves show the phase diagram calculated by the thermodynamic perturbation theory with the modified AO potential. }
\label{fig1}
\end{figure*}

The Helmholtz free energy curves were calculated using the thermodynamic perturbation theory with the effective potential. The phase diagrams were obtained using the common tangent construction on the free energy curves. The phase diagrams for bR monomers (a) and trimers (b) calculated using the AO or the modified AO potentials as the effective potential are shown in FIG. \ref{fig1}.\par
The coexistence region for the modified AO potential expands at a lower lipid packing fraction than the AO potential. Here, we checked the diagram for the monomers with the lipids (see FIG. \ref{fig1} (a)). The broadening for the modified AO potential (solid curve) started at about $\eta_{\rm lip}^{\rm res}  = 0.35$, although that for the AO potential (dashed curve) started at about $\eta_{\rm lip}^{\rm res} = 0.7$. We focused on the phase diagrams around $\eta_{\rm lip}^{\rm res} = 0.5$ because we estimated the lipid packing fraction of a cell membrane as 0.5 in our previous paper\cite{Suda}. The broadenings of the coexistence regions for the modified AO potential started lower than $\eta_{\rm lip}^{\rm res} = 0.5$. On the other hand, those for the AO potential started higher than $\eta_{\rm lip}^{\rm res} = 0.5$ (see FIG. \ref{fig1} (a) and (b)). Therefore, effective potential dependence is critical.\par

\begin{figure}[!t]
\centerline{
\includegraphics[width=8.6cm]{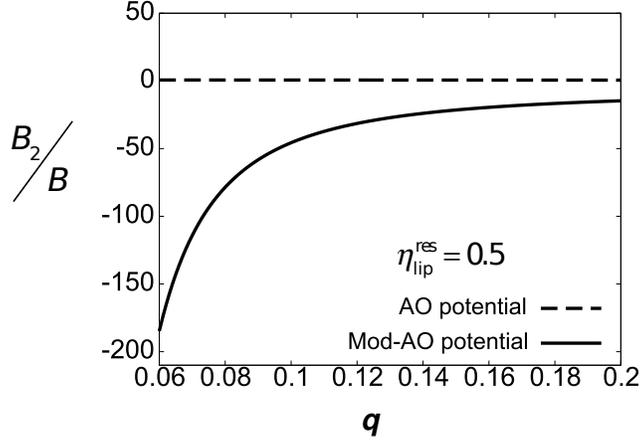}
}
\caption{
The reduced second virial coefficients $B_2/B_2^{HD} $  at $\eta_{\rm lip}^{\rm res} =0.5$. The dashed and the solid curves show the coefficients for the AO and the modified AO potentials, respectively.}
\label{fig2}
\end{figure} 

We defined the critical concentration, CC$(\eta_{\rm lip}^{\rm res})$, as the concentration for the smallest packing fraction for the fluid--ordered phase coexistence state at $\eta_{\rm lip}^{\rm res}$. Thus, the CC($\eta_{\rm lip}^{\rm res}$) gives the boundary curve between the fluid and fluid--ordered states\cite{Suda}. Unfortunately, we could not obtain the phase diagrams for the modified AO potential systems in the region $\eta_{\rm lip}^{\rm res} >0.4$\cite{footnote} (see solid curves in FIG. \ref{fig1}). However, the CC is almost 0 even when $\eta_{\rm lip}^{\rm res} =0.4$. Therefore, we can expect that the CC is very low when $\eta_{\rm lip}^{\rm res} =0.5$. This suggests that the depletion effect can be dominant in the driving force for the bR ordering.\par
The phase diagrams for the AO potential systems give us a contrasting picture (see dashed curves in FIG. \ref{fig1}). Even when $\eta_{\rm lip}^{\rm res} =0.5$, the CC values were almost similar to those for the pure bR system($\eta_{\rm lip}^{\rm res} =0.0$). The ratio of CC value at $\eta_{\rm lip}^{\rm res} =0.5$ to that at $\eta_{\rm lip}^{\rm res} =0.0$ is 0.969 in the case of monomer bRs. The ratio for the trimer bRs is 0.927. Therefore, the depletion effect is weak in the driving force. We obtained the reduced second virial coefficients $\it{B_2/B_2^{\rm HD}}$(FIG. \ref{fig2}). The coefficients for the modified AO potential were negative, and the absolute value was large when $\eta_{\rm lip}^{\rm res} =0.5$. By contrast, those for the AO potential were positive. \par
The direct repulsions between bRs and between bR and lipids were the same when we compared the AO potential with the modified AO potential. Then, the difference in the direct interactions is the core repulsion between lipids. Thus, the core repulsions between lipids caused differences in the coefficients (see FIG. \ref{fig2}) and the phase diagrams (see FIG. \ref{fig1}). When the core repulsions between the lipid molecules exist, the calculated results are consistent with the observation of bR crystallization. Therefore, the core repulsion between the lipid molecules is probably one of the essential factors in the discussion of crystallization.\par

\begin{figure*}[ht]
\centerline{
{\includegraphics[clip, width=8.9cm]{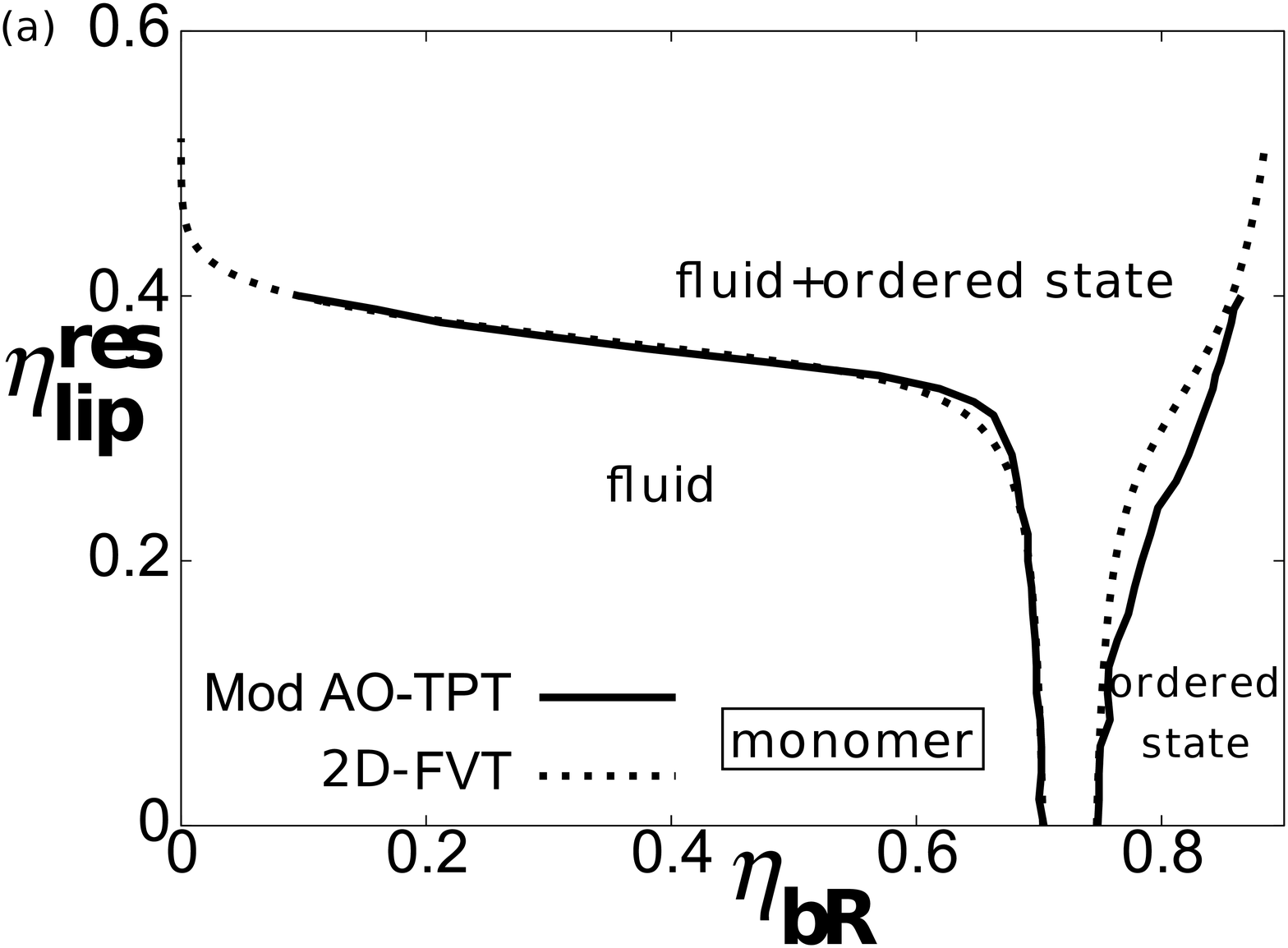}
\label{fig1A}}
\hfil
{\includegraphics[clip, width=8.9cm]{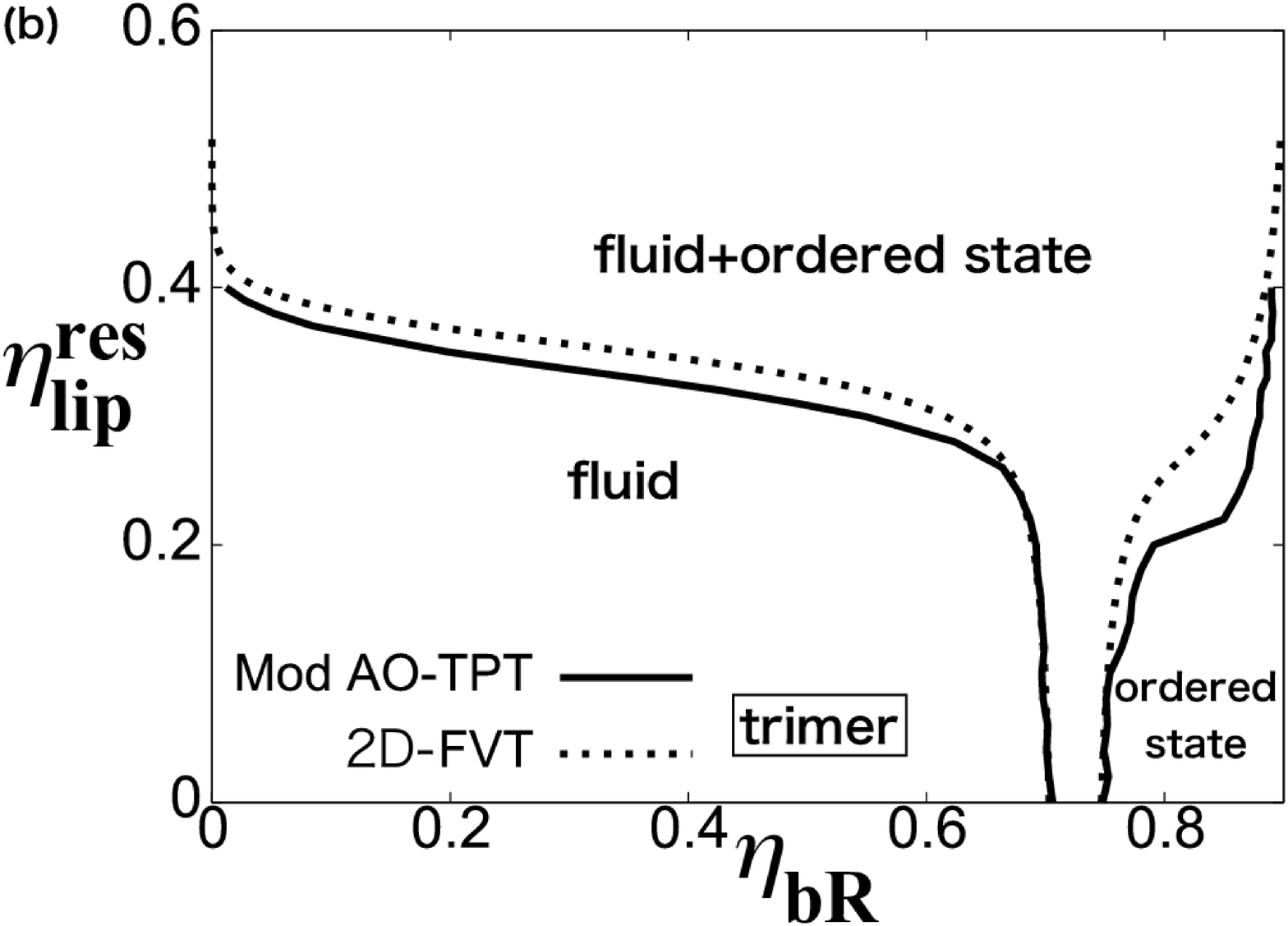}
\label{fig1B}}
}
\caption{Phase diagrams for (a) bR monomer ($q=0.16667$) and (b) trimer ($q=0.08065$). The solid curves show the phase diagrams calculated by the modified AO--TPT approach. The dotted curves show the phase diagrams calculated by the 2D--
FVT approach\cite{Suda}. }
\label{TPTvsFVT}
\end{figure*}

Here, we compared the present study with our previous one\cite{Suda}. In the previous study, the chemical potentials for the fluid and ordered phases were described using the two-dimensional Carnahan-Starling-like equation of state (2D--CSE) and the two-dimensional free volume theory (2D--FVT) in the case of the pure bR system $(\eta_{\rm lip}^{\rm res}=0)$. Thus, the reference systems in both approaches are common. Additionally, we adopted the pressure of the two-dimensional scaled particle theory (2D--SPT) as the reservoir pressure. This means that the repulsive cores of the lipids are common. However, the previous study's theoretical framework did not explicitly contain the effective interaction between two bRs\cite{Suda}. On the other hand, we adopted the modified AO potential as the effective interaction and calculated the phase diagrams using the thermodynamic perturbation theory with the effective interaction (modified AO--TPT) in the present study. The results are compared in FIG. \ref{TPTvsFVT}.\par
Because the theories of the reference systems, namely the pure bR system ($\eta_{\rm lip}^{\rm res}=0$), are common, the coexistence region appears in the same region\cite{RefSys} (see FIG. \ref{TPTvsFVT}). Even when  $\eta_{\rm lip}^{\rm res}$ increases, the phase diagrams of the previous study are in good agreement with the results given by the present modified AO--TPT study. In particular, the boundaries between the fluid and the fluid--ordered coexistence phases in the present study agree very well with the previous results. The previous research shows that the 2D--FVT approach explains the experimental results semi-quantitatively\cite{Suda}. Therefore, we can expect that the modified AO--TPT approach is also valid.\par
On the other hand, we could find a difference between phase diagrams for the bR trimer calculated by modified AO--TPT and 2D--FVT approaches. The coexistence region calculated by the modified AO--TPT approach is slightly wider than that calculated by the 2D--FVT approach. This difference increases as the size ratio \it{q} \rm{approaches} 0 (data is not shown). We cannot deny the problem in the modified AO--TPT approach. However, the results of the 2D--FVT approach are more suspicious because the ratio $q$-dependence on the phase diagram disappears at the small $q$. This problem of the 2D--FVT approach for small $q$ will be explained later. 

\begin{figure*}[t]
\centerline{
{\includegraphics[clip, width=8.9cm]{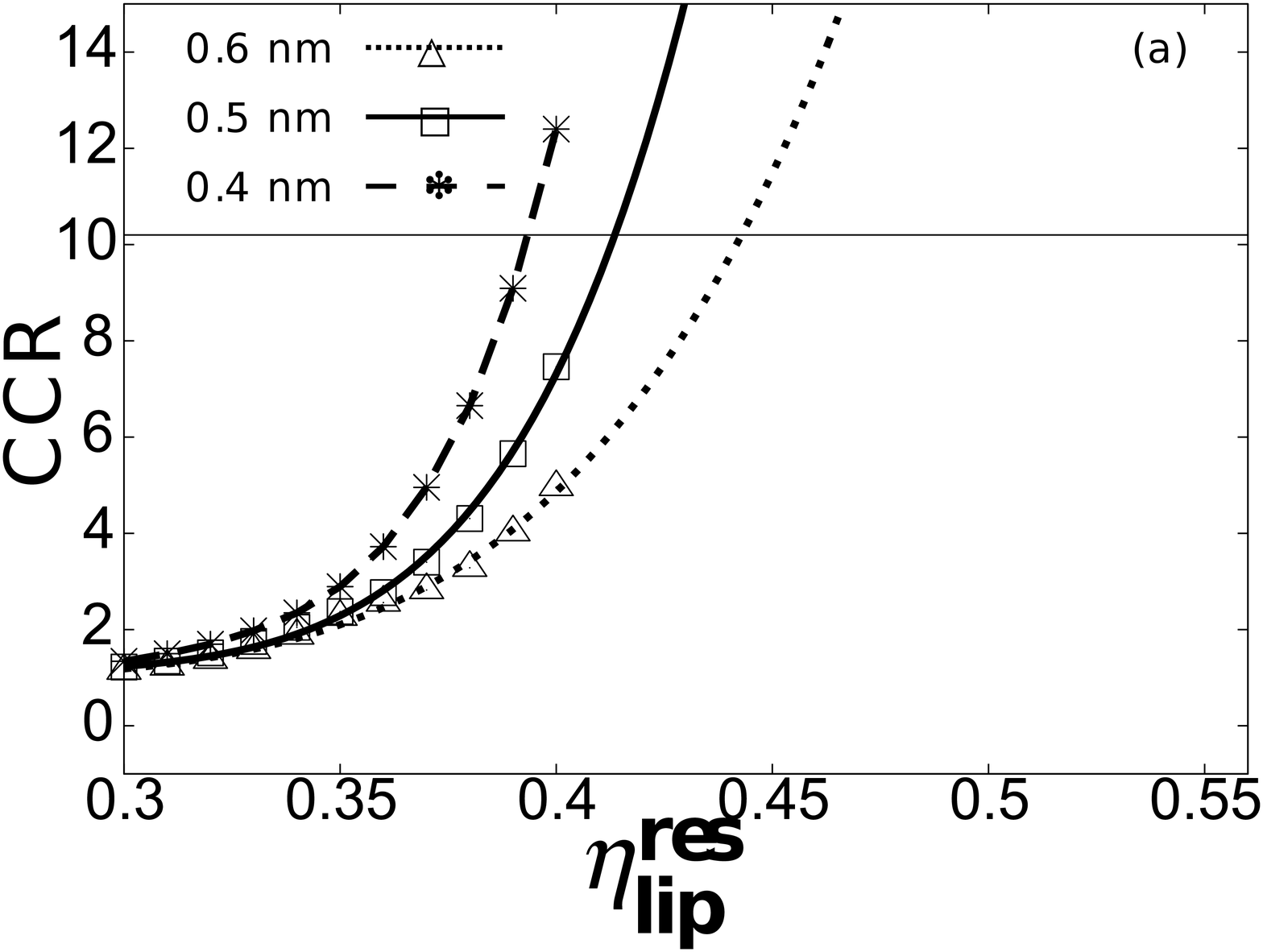}
\label{fig5A}}
\hfil
{\includegraphics[clip, width=8.9cm]{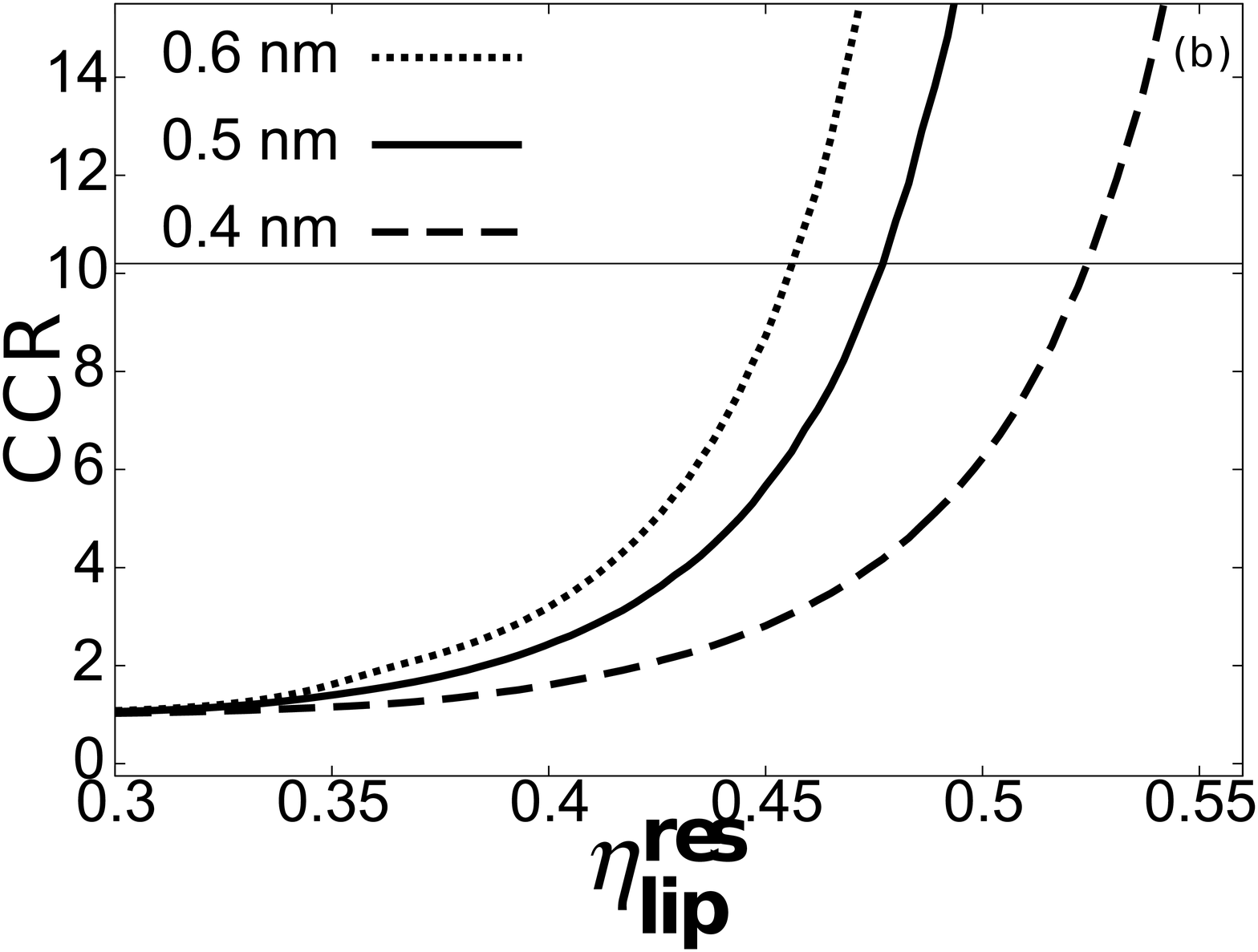}
\label{fig5B}}
}
\caption{(a) The CCRs were calculated by the modified AO--TPT approach. The triangles, squares, and asterisks show the CCR for $\sigma_{\rm lip}=$ 0.4, 0.5, and 0.6, respectively. A dashed line shows the CCR for $\sigma_{\rm lip}=$ 0.4. A solid and dotted lines show the sixth-order polynomial approximation curves of CCR for $\sigma_{\rm lip}=$ 0.5 and 0.6, respectively. The experimental value of the CCR, 10.2, is shown as a thin solid line parallel to the x--axis. (b) The CCRs were calculated by the 2D--FVT approach. The dotted line, a solid line, and a dashed line show the CCR for $\sigma_{\rm lip}=$ 0.4, 0.5, and 0.6, respectively\cite{Suda}. }
\label{CCR}
\end{figure*}

The CC for the bR trimer ordering is lower than for the monomer ordering in both theories. For example, when $\eta_{\rm lip}^{\rm res} =0.4$, the CCs for bR monomer ordering and trimer ordering calculated by the 2D--FVT approach are 0.100 and 0.042, respectively\cite{Suda}. On the other hand, the CCs for the bR monomer ordering and the trimer ordering calculated by the modified AO--TPT approach are 0.097 and 0.013, respectively. The CC for the bR trimer ordering is lower than that for the monomer ordering in each theory. These results correspond to experimental results\cite{CCR} qualitatively. As with the 2D--FVT approach\cite{Suda}, the modified AO--TPT approach explains the experimental results for bR crystallization\cite{CCR} qualitatively.\par
FIG. \ref{CCR} (a) shows the critical concentration ratios (CCRs) between bR trimer and monomer obtained from the phase diagrams calculated using the modified AO--TPT approach (symbol). The curves are drawn by sixth-order polynomial approximation when $\sigma_{\rm lip} =$ 0.5 and 0.6 nm. On the other hand, the polynomial approximation for $\sigma_{\rm lip} =$ 0.4 is not drawn because the data are across the line for the CCR=10.2. FIG. \ref{CCR} (b) has the CCR plots previously calculated by the 2D--FVT approach\cite{Suda} for comparison. In the present and previous studies\cite{Suda}, the diameter of the lipid molecule is 0.5 nm as a standard. However, this model has arbitrariness. The three lipid diameters $\sigma_{\rm lip} =$ 0.4, 0.5, and 0.6 nm, were examined to remove the arbitrariness. The experimental value of the CCR, 10.2, is shown as a solid line parallel to the x--axis. When the CC is too low, we cannot obtain the phase diagram using the TPT approach with the modified AO interaction because the common tangent cannot be drawn on the free energy curves. Therefore, the CCR cannot be calculated up to 10.2. Thus, the extrapolated plots were drawn when $\sigma_{\rm lip} =$ 0.5 and 0.6 nm. The CCR increases in each theory as $\eta_{\rm lip}^{\rm res}$ increases.\par
The CCRs obtained by the modified AO--TPT approach and the 2D--FVT approach were compared. In the modified AO--TPT approach, the maximum $\eta_{\rm lip}^{\rm res}$ where CC can be calculated for all lipid diameters was 0.40. At $\eta_{\rm lip}^{\rm res}=$0.40, the CCRs obtained by the two approaches were compared. When the $\sigma_{\rm lip}$ is 0.4 nm, the CCR obtained by the modified AO--TPT approach was 12.4. On the other hand, the CCR obtained by the 2D--FVT approach was 1.60. When the $\sigma_{\rm lip}$ is 0.5 nm, the CCR obtained by the modified AO--TPT approach was 7.46. On the other hand, the CCR obtained by the 2D--FVT approach was 2.43. When the $\sigma_{\rm lip}$ is 0.6 nm, the CCR obtained by the modified AO--TPT approach was 4.96. On the other hand, the CCR obtained by the 2D--FVT approach was 3.18. The CCR obtained by the modified AO--TPT approach was larger than that obtained by the 2D--FVT approach for all $\sigma_{\rm lip}$. This is because the CC for trimer calculated by the modified AO--TPT approach is smaller than that calculated by the 2D--FVT approach, while the CC for monomer is almost the same between the two theories.\par
The extrapolated CCR plots obtained by the modified AO--TPT approach agree with experimental CCR value 10.2 at $\eta_{\rm lip}^{\rm res}=$ 0.393, 0.414, 0.443 for $\sigma_{\rm lip} =$ 0.4, 0.5, 0.6 nm, respectively. $\eta_{\rm lip}^{\rm res}$s correspond to the lipid number densities 3.13, 2.11, and 1.57 $\times 10^6 \mu m^{-2}$, respectively. Furthermore, the CCRs obtained by the modified 2D--FVT approach agree with the experimental results at the lipid number density of 4.17, 2.43, 1.61 $\times 10^6 \mu m^{-2}$ for $\sigma_{\rm lip} =$ 0.4, 0.5, 0.6 nm, respectively\cite{Suda}. On the other hand, the lipid number density in the single layer of a cell membrane was about 2.5 $\times 10^6 \mu m^{-2}\ $\cite{cell}. Therefore, we expect the CCRs to almost agree with experimental results if we consider the lipid molecule's repulsive core.\par
The phase diagrams and CCRs for the 2D--FVT and the modified AO--TPT approaches are almost identical. On the other hand, we can find a qualitative difference between their CCRs. Comparing FIG. \ref{CCR} (a) and (b), the CCRs order of the dependence on the lipid diameter for the modified AO--TPT approach is the inverse of that for the 2D--FVT approach. The CCRs obtained by the 2D--FVT approach increase as the lipid diameter increases. In contrast, the CCRs obtained by the modified AO--TPT approach decrease as the lipid diameter increases. This discrepancy seems to be caused by a problem in the 2D--FVT approach: namely, the calculation of CC for the trimer. García et al. reported that the $q$-dependence of the phase diagram obtained by the FVT approach disappears in a 3D system as the value $q$ decreases\cite{GEO}. They further noticed that this result did not agree with an experimental result\cite{GEO}. This problem in the 3D systems should be conserved in the 2D systems.\par
\begin{figure}[!t]
\centerline{
\includegraphics[width=8.6cm]{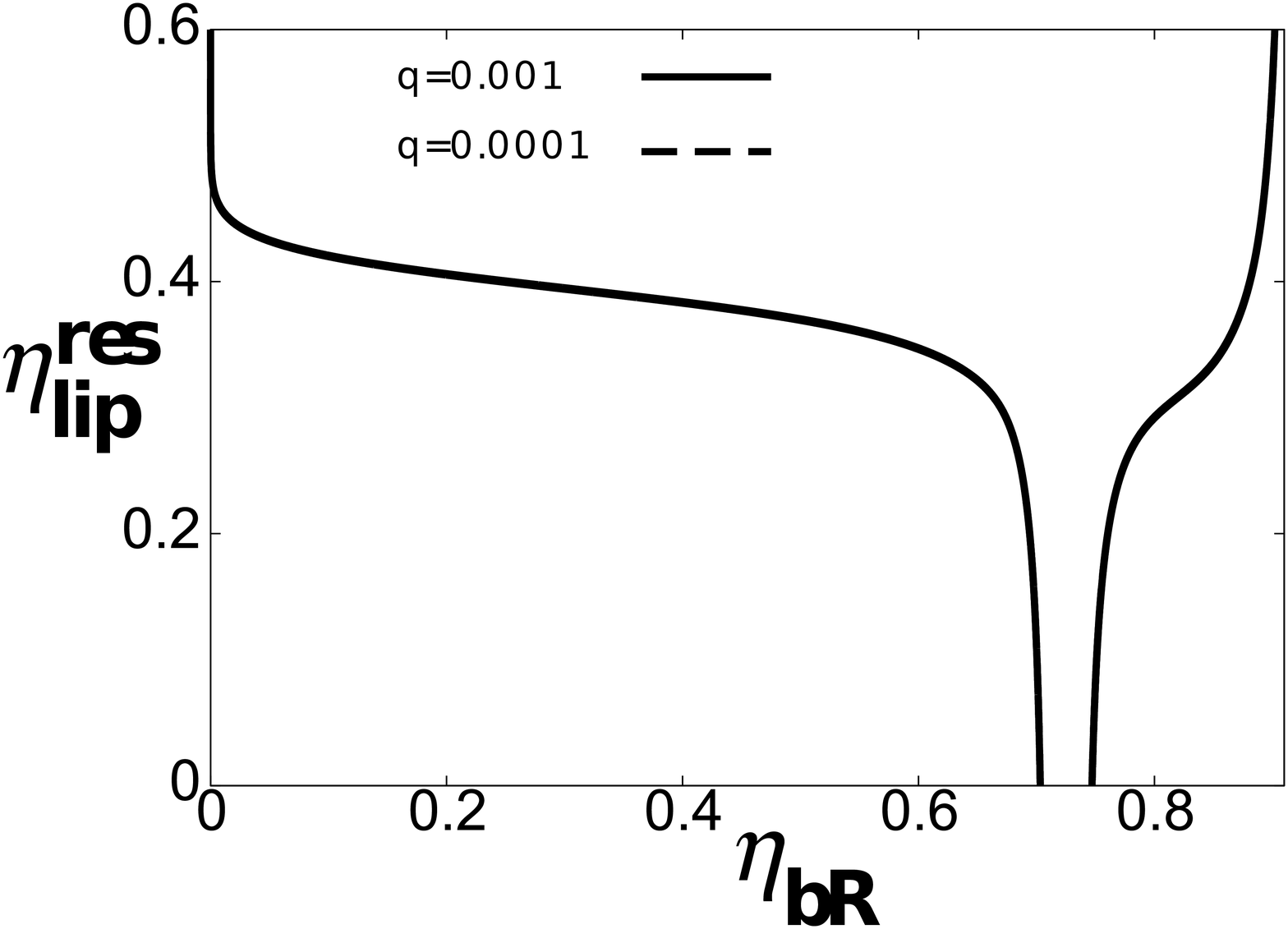}
}
\caption{
The phase diagrams for small $q$s calculated by the 2D--FVT approach. The solid and dashed curves show the phase diagrams for $q=$0.001 and 0.0001, respectively. These coexistence regions almost overlap each other.
}
\label{smallq}
\end{figure}

We examined the $q$-dependence using the 2D--FVT approach. FIG. \ref{smallq} shows the phase diagrams for $q =$ 0.001 and 0.0001 calculated using the 2D--FVT approach. These coexistence regions almost overlap each other. This agreement numerically shows that the theoretical approach does not have $q$ dependence when $q$ is small. Thus, we can discuss the relationship between CCR and lipid diameter in the 2D--FVT approach, as follows. The dependence on the lipid diameter for bR monomer CC is more significant than that for bR trimer CC because the value $q$ for monomer CC (about 0.16667) is more significant than that for trimer CC (about 0.08065). Here, the dependence on the lipid diameter is also the $q$-dependence. Therefore, the CCR dependence mainly obeys the dependence on the lipid diameter for the monomers, and the CCR increases as the lipid diameter increases in the case of the 2D--FVT approach.\par

The disappearance of the $q$-dependence for the 2D--FVT approach at small $q$ has to be explained. In the phase diagram calculation, we must find the bR packing fraction for the fluid phase $\eta_{\rm bR}^{\rm fluid}$ and that for the ordered phase, $\eta_{\rm bR}^{\rm ordered}$. We obtained them to solve the equations (\ref{p}) and (\ref{mu})\cite{Suda,Lekkerkerker}. The pressure and the chemical potential are expressed by 2D--FVT\cite{Suda,Lekkerkerker}, as follows:
\begin{eqnarray}
\beta p_{\rm f} v_{\rm bR}= \beta p^0_{\rm f} v_{\rm bR}+\beta p^{\rm res}v_{\rm bR}\left( \alpha_{\rm f}-\eta_{\rm bR}^{\rm fluid}\left(\frac{\partial \alpha_{\rm f}}{\partial \eta_{\rm bR}^{\rm fluid}} \right)_{N_{\rm bR},T,\mu_{\rm lip}}\right),\label{pf}\\
\beta p_{\rm ord} v_{\rm bR}=\beta p^0_{\rm ord} v_{\rm bR}+\beta p^{\rm res}v_{\rm bR}\left( \alpha_{\rm ord}-\eta_{\rm bR}^{\rm ordered}\left(\frac{\partial \alpha_{\rm ord}}{\partial \eta_{\rm bR}^{\rm ordered}} \right)_{N_{\rm bR},T,\mu_{\rm lip}}\right),\label{pord}
 \end{eqnarray}
 \begin{eqnarray}
 \beta \mu_{\rm f, bR}&=&\beta \mu_{\rm f, bR}^0-\beta p^{\rm res}v_{\rm bR}\left(\frac{\partial \alpha_{\rm f} }{\partial \eta_{\rm bR}^{\rm fluid} } \right)_{V,T,\mu_{\rm lip}}\label{mufl}, \\
 \beta \mu_{\rm ord, bR}&=&\beta \mu_{\rm ord, bR}^0-\beta p^{\rm res}v_{\rm bR}\left(\frac{\partial \alpha_{\rm ord} }{\partial \eta_{\rm bR}^{\rm ordered} } \right)_{V,T,\mu_{\rm lip}}\label{muord},
\end{eqnarray} 
where, $\alpha_{\rm f}$ and $\alpha_{\rm ord}$ are the free volume fraction for the fluid phase and ordered phase, respectively. $p_{\rm f}^0$ and $p_{\rm ord}^0$ are the pressure for the fluid and ordered phases in a bR pure system. $\mu_{\rm f,bR}^0$ and $\mu_{\rm ord,bR}^0$ are the chemical potential for the fluid and ordered phases in the bR pure system. If these values are substituted to the relations (\ref{p}) and (\ref{mu}), namely $p_{\rm f}=p_{\rm ord}$, $\mu_{\rm f,bR}=\mu_{\rm ord,bR}$, we obtain two equations, as follows:
\begin{equation}
\beta \left( p^0_{\rm f}- p^0_{\rm ord}\right) v_{\rm bR}+\beta p^{\rm res}v_{\rm bR}\left( \alpha_{\rm f}-\eta_{\rm bR}^{\rm fluid}\left(\frac{\partial \alpha_{\rm f}}{\partial \eta_{\rm bR}^{\rm fluid}} \right) - \alpha_{\rm ord}+\eta_{\rm bR}^{\rm ordered}\left(\frac{\partial \alpha_{\rm ord}}{\partial \eta_{\rm bR}^{\rm ordrered}} \right) \right)=0\label{p-p},
 \end{equation} 

\begin{equation}
    \beta\left(\mu_{\rm f, bR}^0-\mu_{\rm ord, bR}^0 \right)-\beta p^{\rm res}v_{\rm bR}\left(\left(\frac{\partial \alpha_{\rm f} }{\partial \eta_{\rm bR}^{\rm fluid} }\right)-\left(\frac{\partial \alpha_{\rm ord} }{\partial \eta_{\rm bR}^{\rm ordered} }\right)\right)=0\label{mu-mu}.
\end{equation}
$\alpha_{\rm f}$ and $\alpha_{\rm ord}$ are obtained by SPT\cite{Suda,Lekkerkerker} as follows:

\begin{eqnarray}
\alpha_{\rm f}&=&\left( 1-\eta_{\rm bR}^{\rm fluid}  \right){\rm exp}\left[- \frac{2\eta_{\rm bR}^{\rm fluid} q}{1- \eta_{\rm bR}^{\rm fluid} }  -\frac{\eta_{\rm bR}^{\rm fluid} q^2 }{\left(1-\eta_{\rm bR}^{\rm fluid} \right)^2} \right],\\
\alpha_{\rm ord}&=&\left( 1-\eta_{\rm bR}^{\rm ordered}  \right){\rm exp}\left[- \frac{2\eta_{\rm bR}^{\rm ordered} q}{1- \eta_{\rm bR}^{\rm ordered} }  -\frac{\eta_{\rm bR}^{\rm ordered} q^2 }{\left(1-\eta_{\rm bR}^{\rm ordered} \right)^2} \right].
\end{eqnarray}
The partial differentiations are 
\begin{eqnarray}
\left( \frac{\partial \alpha_{\rm f}}{\partial \eta_{\rm bR}^{\rm fluid}}\right)=&\frac{-(\eta_{\rm bR}^{\rm fluid})^2 +(-q^2+2q+2)\eta_{\rm bR}^{\rm fluid} -q^2-2q-1}{(1-\eta_{\rm bR}^{\rm fluid})^2}\nonumber\nonumber\\
& {\rm exp}\left[- \frac{2\eta_{\rm bR}^{\rm fluid} q}{1- \eta_{\rm bR}^{\rm fluid} }  -\frac{\eta_{\rm bR}^{\rm fluid} q^2 }{\left(1-\eta_{\rm bR}^{\rm fluid} \right)^2} \right],
\end{eqnarray}

\begin{eqnarray}
\left( \frac{\partial \alpha_{\rm ord}}{\partial \eta_{\rm bR}^{\rm ordered}}\right)=&\frac{-(\eta_{\rm bR}^{\rm ordered})^2 +(-q^2+2q+2)\eta_{\rm bR}^{\rm ordered} -q^2-2q-1}{(1-\eta_{\rm bR}^{\rm ordered})^2}\nonumber\\
& {\rm exp}\left[- \frac{2\eta_{\rm bR}^{\rm ordered} q}{1- \eta_{\rm bR}^{\rm ordered} }  -\frac{\eta_{\rm bR}^{\rm ordered} q^2 }{\left(1-\eta_{\rm bR}^{\rm ordered} \right)^2} \right].
\end{eqnarray}
The pressure in the reservoir is expressed using SPT\cite{Suda,Lekkerkerker}: namely, eq. (\ref{pres}). Then, $\beta p^{\rm res}v_{\rm bR}$ is expressed using $\eta_{\rm lip}^{\rm res}$ as, follows:

\begin{equation}
\beta p^{\rm res} v_{\rm bR}=\frac{\eta_{\rm lip}^{\rm res}q^{-2}}{\left(1-\eta_{\rm lip}^{\rm res} \right)^2}.
\label{}
\end{equation}

The first term on the left-hand side in eqs. (\ref{p-p}) and (\ref{mu-mu}) does not depend on $q$. On the other hand, we must confirm the behavior of the second term on the left-hand side in eq. (\ref{p-p}) at the limit of $¥q = 0$ because the part $\beta p^{\rm res} v_{\rm bR}$ goes to infinity, and the remaining part---namely, the part in the parentheses---goes to zero. The second term on the left-hand side in eq. (\ref{mu-mu}) also has a similar problem. Thus, these second terms are expanded to polynomials to discuss the $q$-dependence.\par
The left-hand sides of eqs. (\ref{p-p}) and (\ref{mu-mu}) are expanded to the polynomial functions of $q$, as follows\cite{SupMat}.

\begin{equation}
(p^0_{\rm f}-p^0_{\rm ord})v_{\rm bR}+\frac{\eta_{\rm lip}^{\rm res}}{(1-\eta_{\rm lip}^{\rm res})^2}\left( \frac{-\left( \eta_{\rm bR}^{\rm fluid}\right)^2}{(1-\eta_{\rm bR}^{\rm fluid})^2}+\frac{\left( \eta_{\rm bR}^{\rm ordered}\right)^2}{(1-\eta_{\rm bR}^{\rm ordered})^2} \right)+O(q)=0
\label{p-p0}
\end{equation}

\begin{eqnarray}
&&\mu^0_{\rm f}-\mu_{\rm ord}^0+\nonumber\\
&&\frac{\eta_{\rm lip}^{\rm res}}{(1-\eta_{\rm lip}^{\rm res})^2} \left( \frac{1-4\eta_{\rm bR}^{\rm fluid}+2\left(\eta_{\rm bR}^{\rm fluid}\right)^2}{(1-\eta_{\rm bR}^{\rm fluid})^2}-\frac{1-4\eta_{\rm bR}^{\rm ordered}+2\left(\eta_{\rm bR}^{\rm ordered}\right)^2}{(1-\eta_{\rm bR}^{\rm ordered})^2} \right)\nonumber\\
&&+O(q)=0
\label{mu-mu0}
\end{eqnarray}
These limit equations gave us a phase diagram that almost overlaped those in FIG. \ref{smallq}. The $q-$dependence disappears at the limit of $q=0$. Therefore, the phase diagram of the 2D--FVT approach does not depend on $q$ for small $q$. \par
This disappearance of the $q$-dependence originates in $\alpha_{\rm f}$ and $\alpha_{\rm ord}$. The same theory, namely SPT, is used to express both $\alpha_{\rm f}$ and $\alpha_{\rm ord}$. Therefore, the 0th-order terms of $q$ for the fluid and ordered phases cancel each other in eqs. (\ref{p-p0}) and (\ref{mu-mu0}). In the 3D system, a similar problem has been pointed out, and the geometrical consideration for $\alpha_{\rm ord}$ was proposed instead of SPT\cite{GEO,remco2021}.
\section{CONCLUSION}
The phase diagrams of binary hard-disk systems were calculated to discuss the driving force of bR crystallization. To study the effects of the core repulsive force between lipid molecules as depletants, we calculated the phase diagram and CCR using the thermodynamic perturbation theory with two effective interactions between bRs\cite{Suematsu2016,Velasco,TP}. The current approach was entirely different from the previous study based on simple theories \cite{Suda}. However, the results agreed with the experimental ones only when we considered the core repulsive force between lipid molecules. Despite the different calculation methods, the results were the same as the previous one. Therefore, the present results also support that the core repulsion between lipid molecules plays an essential role in the driving force of crystallization. In the future, we think that verification by simulation should be necessary.
\section{Acknowledgments}
This work was also supported by JapanSociety for the Promotion of Science (JSPS) KAKENHI Grant Nos. JP22K03558 and JP21K18604, JP19H01863, JP19K03772, and JP22J13118. The computation was performed using Research Center for Computational Science, Okazaki, Japan (Project: 22-IMS-C101 and 22-IMS-C113) and the Research Institute for Information Technology, Kyushu University.
\bibliographystyle{elsarticle-num} 
 \bibliography{cas-refs}

\providecommand{\noopsort}[1]{}\providecommand{\singleletter}[1]{#1}%
\begin{thebibliography}{10}
\expandafter\ifx\csname url\endcsname\relax
  \def\url#1{\texttt{#1}}\fi
\expandafter\ifx\csname urlprefix\endcsname\relax\def\urlprefix{URL }\fi
\expandafter\ifx\csname href\endcsname\relax
  \def\href#1#2{#2} \def\path#1{#1}\fi

\bibitem{RHenderson}
R.~Henderson, P.~N.~T. Unwin, Nature 257 (1975) 28, \\
  https://doi.org/10.1038/257028a0.

\bibitem{RHenderson2}
R.~Henderson, J.~M. Baldwin, T.~A. Ceska, F.~Zemlin, E.~Beckmann, K.~H.
  Downing, J.\ Mol.\ Biol. 213 (1990) 899, \\
  https://doi.org/10.1016/S0022-2836(05)80271-2.

\bibitem{review}
M.~P. Krebs, T.~A. Isenbarger, Biochim.\ Biophys.\ Acta 1460 (2000) 15, \\
  https://doi.org/10.1016/S0005-2728(00)00126-2.

\bibitem{Suda}
K.~Suda, A.~Suematsu, R.~Akiyama, J.\ Chem.\ Phys. 154 (2021) 204904, \\
  https://doi.org/10.1063/5.0044399.

\bibitem{AO}
S.~Asakura, F.~Oosawa, J.\ Chem.\ Phys. 22 (1954) 1255, \\
  https://doi.org/10.1063/1.1740347.

\bibitem{AOpolymer}
S.~Asakura, F.~Oosawa, J. Pol. Sci. 33 (1958) 183, \\
  https://doi.org/10.1002/pol.1958.1203312618.

\bibitem{FMT1}
R.~Roth, B.~G{\"o}tzelmann, S.~Dietrich, Phys. Rev. Lett. 83 (1999) 448, \\
  https://doi.org/10.1103/PhysRevLett.83.448.

\bibitem{FMT2}
Y.~Rosenfeld, Phys. Rev. Lett. 63 (1989) 980, \\
  https://doi.org/10.1103/PhysRevLett.63.980.

\bibitem{FMT3}
R.~Roth, R.~Evans, A.~Lang, G.~Kahl, J. Phys.: Condens. Matter 14 (2002) 12063,
  \\ https://doi.org/10.1088/0953-8984/14/46/313.

\bibitem{NakamuraAkiyama}
Y.~Nakamura, S.~Arai, M.~Kinoshita, A.~Yoshimori, R.~Akiyama, J. Chem. Phys.
  151 (2019) 044506, \\ https://doi.org/10.1063/1.5100040.

\bibitem{vrij}
A.~Vrij, Pure Appl. Chem. 48 (1976) 471, \\
  https://doi.org/10.1351/pac197648040471.

\bibitem{Lekkerkerker}
H.~N.~W. Lekkerkerker, R.~Tuinier, {\it Colloids and the Depletion
  Interaction}, Springer, New York, 2011, \\
  htpps://doi.org/10.1007/978-94-007-1223-2.

\bibitem{Lekkerkerker2}
H.~N.~W. Lekkerkerker, W.~C.~K. Poon, P.~N. Pusey, A.~Stroobants, P.~B. Warren,
  Europhys. Lett. 20 (1992) 559, \\ https://doi.org/10.1209/0295-5075/20/6/015.

\bibitem{Frenkel}
D.~Frenkel, A.~A. Louis, Phys. Rev. Lett. 68 (1992) 3363, \\
  https://doi.org/10.1103/PhysRevLett.68.3363.

\bibitem{Dijkstra1}
M.~Dijkstra, D.~Frenkel, J.~P. Hansen, J. Chem. Phys. 101 (1994) 3179,
  \\https://doi.org/10.1063/1.468468.

\bibitem{Dijkstra2}
M.~Dijkstra, D.~Frenkel, Phys. Rev. Lett. 72 (1994) 298, \\
  https://doi.org/10.1103/PhysRevLett.72.298.

\bibitem{Dijkstra3}
{M. Dijkstra and R. van Roij and R. Evans}, Phys. Rev. Lett. 81 (1998) 2268, \\
  https://doi.org/10.1103/PhysRevLett.81.2268.

\bibitem{Dijkstra4}
{M. Dijkstra and R. van Roij and R. Evans}, Phys. Rev. Lett. 82 (1999) 117, \\
  https://doi.org/10.1103/PhysRevLett.82.117.

\bibitem{Dijkstra5}
{M. Dijkstra and R. van Roij and R. Evans}, Phys. Rev. E 59 (1999) 5744, \\
  https://doi.org/10.1103/PhysRevE.59.5744.

\bibitem{Dijkstra6}
M.~Dijkstra, J.~M. Brader, R.~Evans, J. Phys.: Condens. Matter 11 (1999) 10079,
  \\ https://doi.org/10.1088/0953-8984/11/50/304.

\bibitem{Dijkstra7}
{M. Dijkstra and R. van Roij and R. Roth and A. Fortini}, Phys. Rev. E 73
  (2006) 041404, \\ https://doi.org/10.1103/PhysRevE.73.041404.

\bibitem{Cinacchi}
G.~Cinacchi, Y.~Mart{\'i}nez-Rat{\'o}n, L.~Mederos, G.~Navascu{\'e}s, A.~Tani,
  E.~Velasco, J. Chem. Phys. 127 (2007) 214501, \\
  https://doi.org/10.1063/1.2804330.

\bibitem{Lopez}
E.~L{\'o}pez-S{\'a}nchez, C.~D. Estrada-{\'A}lvarez, G.~P{\'e}rez-{\'A}ngel,
  J.~M. M{\'e}ndez-Alcaraz, P.~Gonz{\'a}lez-Mozuelos, R.~Castañeda-Priego, J.
  Chem. Phys. 139 (2013) 104908, \\ https://doi.org/10.1063/1.4820559.

\bibitem{Velasco}
E.~Velasco, G.~Navascu{\'e}s, L.~Mederos, Phys. Rev. E 60 (1999) 3158, \\
  https://doi.org/10.1103/PhysRevE.60.3158.

\bibitem{Suematsu2016}
A.~Suematsu, A.~Yoshimori, R.~Akiyama, Europhys. Lett. 116 (2016) 38004, \\
  https://doi.org/10.1209/0295-5075/116/38004.

\bibitem{Chang}
A.~M. Tom, W.~K. Kim, C.~Hyeon, J. Chem. Phys. 154 (2021) 214901, \\
  https://doi.org/10.1063/5.0048554.

\bibitem{monomer}
M.~P. Krebs, W.~Li, T.~P. Halambeck, J.\ Mol.\ Biol. 267 (1997) 172, \\
  https://doi.org/10.1006/jmbi.1996.0848.

\bibitem{CCR}
T.~A. Isenbarger, M.~P. Krebs, Biochemistry 38 (1999) 9023, \\
  https://doi.org/10.1021/bi9905563.

\bibitem{Rovert}
J.~T. Lee, M.~Robert, Phys. Rev. E 60 (1999) 7198, \\
  https://doi.org/10.1103/PhysRevE.60.7198.

\bibitem{disappear}
J.~Russo, N.~B. Wilding, Phys.\ Rev.\ Lett. 119 (2017) 115702, \\
  https://doi.org/10.1103/PhysRevLett.119.115702.

\bibitem{Ott}
S.-C. Lin, M.~Oettel, Phys. Rev. E 98 (2018) 012608, \\
  https://doi.org/10.1103/PhysRevE.98.012608.

\bibitem{SPT}
H.~Reiss, H.~L. Frisch, J.~L. Lebowitz, J.\ Chem.\ Phys. 31 (1959) 369, \\
  https://doi.org/10.1063/1.1730361.

\bibitem{2DSPT}
E.~Helfand, H.~L. Frisch, J.~L. Lebowitz, J.\ Chem.\ Phys. 34 (1961) 1037, \\
  https://doi.org/10.1063/1.1731629.

\bibitem{TP}
Y.~Tamura, A.~Yoshimori, A.~Suematsu, R.~Akiyama, Europhys. Lett 129 (2020)
  66001, \\ https://doi.org/10.1209/0295-5075/129/66001.

\bibitem{6.2}
A.~Rakovich, A.~Sukhanova, N.~Bouchonville, E.~Lukashev, V.~Oleinikov,
  M.~Artemyev, V.~Lesnyak, N.~Gaponik, M.~Molinari, M.~Troyon, Y.~P. Rakovich,
  J.~F. Donegan, I.~Nabiev, Nano\ Lett. 10 (2010) 2640, \\
  https://doi.org/10.1021/nl1013772.

\bibitem{cell}
B.~Alberts, A.~Johnson, J.~Lewis, D.~Morgan, M.~Raff, K.~Roberts, P.~Walter,
  {\it Molecular Biology of THE CELL}, Garland Science, New York, 2015.

\bibitem{hexatic}
E.~P. Bernard, W.~Krauth, Phys.\ Rev.\ Lett. 107 (2011) 155704, \\
  https://doi.org/10.1103/PhysRevLett.107.155704.

\bibitem{2DFVT}
Z.~W. Salsburg, W.~W. Wood, J.\ Chem.\ Phys. 37 (1962) 798, \\
  https://doi.org/10.1063/1.1733163.

\bibitem{footnote}
The phase diagram cannot be calculated by modified AO--TPT approach when the CC
  is very low because the common tangent cannot be drawn on the free energy.

\bibitem{RefSys}
The coexistence regions are $\eta_{\rm bR}=0.703-0.747$(2D--FVT approach) and
  $\eta_{\rm bR}=0.703-0.746$(TPT approach). There is slight difference because
  the calculation methods are different.

\bibitem{GEO}
{\'A}.~G. Garc{\'i}a, J.~Opdam, R.~Tuinier, M.~Vis, Chem. Phys. Lett. 709
  (2018) 16, \\ https://doi.org/10.1016/j.cplett.2018.08.028.

\bibitem{SupMat}
See the supplementary material for the derivation.

\bibitem{remco2021}
J.~Opdam, M.~P.~M. Schelling, R.~Tuinier, J. chem. Phys. 154 (2021) 074902, \\
  https://doi.org/10.1063/5.0037963.

\end{thebibliography}


Not Found
\newpage
\vspace*{12pt}
{\noindent \large Supplementary Material for Two-Dimensional Ordering of Bacteriorhodopsins in a Lipid Bilayer and Effects Caused by Repulsive Core between Lipid Molecules on Lateral Depletion Interaction: A Study based on a Thermodynamic Perturbation Theory}
\vskip\baselineskip
The chemical potential for bR and the pressure in the fluid and ordered phases, $\mu_{\rm f, bR}$, $\mu_{\rm ord, bR}$, $p_{\rm f}$,  and $p_{\rm ord}$ are obtained by 2D--FVT as follows:
\begin{eqnarray}
 \beta \mu_{\rm f, bR}&=&\beta \mu_{\rm f, bR}^0-\beta p^{\rm res}v_{\rm bR}\left(\frac{\partial \alpha_{\rm f} }{\partial \eta_{\rm bR}^{\rm fluid} } \right)_{V,T,\mu_{\rm lip}}\label{Smufl},\\
 \beta \mu_{\rm ord, bR}&=&\beta \mu_{\rm ord, bR}^0-\beta p^{\rm res}v_{\rm bR}\left(\frac{\partial \alpha_{\rm ord} }{\partial \eta_{\rm bR}^{\rm ordered} } \right)_{V,T,\mu_{\rm lip}}\label{Smuord},
 \end{eqnarray} 
 \begin{equation}
 \beta p_{\rm f} v_{\rm bR}= \beta p^0_{\rm f} v_{\rm bR}+\beta p^{\rm res}v_{\rm bR}\left( \alpha_{\rm f}-\eta_{\rm bR}^{\rm fluid}\left(\frac{\partial \alpha_{\rm f}}{\partial \eta_{\rm bR}^{\rm fluid}} \right)_{N_{\rm bR},T,\mu_{\rm lip}}\right),\label{Spf}
 \end{equation}
 \begin{equation}
 \beta p_{\rm ord} v_{\rm bR}=\beta p^0_{\rm ord} v_{\rm bR}+
 \beta p^{\rm res}v_{\rm bR}\left( \alpha_{\rm ord}-\eta_{\rm bR}^{\rm ordered}\left(\frac{\partial \alpha_{\rm ord}}{\partial \eta_{\rm bR}^{\rm ordered}} \right)_{N_{\rm bR},T,\mu_{\rm lip}}\right),\label{Spord}
\end{equation} 
where $\beta$ is the inverse temperature, $\mu_{\rm f}^0$ and $\mu_{\rm ord}^0$ are the chemical potential for the bR pure system in the fluid and ordered phases, $p_{\rm f}^0$ and $p_{\rm ord}^0$ are the pressure for the bR pure system in the fluid and ordered phases, $v_{\rm bR}$ is the area of one bR, $p^{\rm res}$ is the pressure in the reservoir, $\alpha_{\rm f}$ and $\alpha_{\rm ord}$ are the free volume fraction in the fluid and ordered phases, and $\eta_{\rm bR}^{\rm fluid}$ and $\eta_{\rm bR}^{\rm ordred}$ are the packing fraction of bRs in the fluid and ordered phases. The phase diagrams were obtained by 2D--FVT approach in the previous study\cite{Suda} solving the following two equations:
\begin{equation}
    \beta\left(\mu_{\rm f, bR}^0-\mu_{\rm ord, bR}^0 \right)-\beta p^{\rm res}v_{\rm bR}\left(\left(\frac{\partial \alpha_{\rm f} }{\partial \eta_{\rm bR}^{\rm fluid} }\right)-\left(\frac{\partial \alpha_{\rm ord} }{\partial \eta_{\rm bR}^{\rm ordered} }\right)\right)=0\label{Smu-mu},
\end{equation}
\begin{equation}
\beta \left( p^0_{\rm f}- p^0_{\rm ord}\right) v_{\rm bR}+\beta p^{\rm res}v_{\rm bR}\left( \alpha_{\rm f}-\eta_{\rm bR}^{\rm fluid}\left(\frac{\partial \alpha_{\rm f}}{\partial \eta_{\rm bR}^{\rm fluid}} \right) - \alpha_{\rm ord}+\eta_{\rm bR}^{\rm ordered}\left(\frac{\partial \alpha_{\rm ord}}{\partial \eta_{\rm bR}^{\rm ordrered}} \right) \right)=0\label{Sp-p}.
 \end{equation} 
$\mu_{\rm f}^0$, $\mu_{\rm ord}^0$, $p_{\rm f}^0$, and $p_{\rm ord}^0$ are independent of $q$. The pressure for the reservoir, $p^{\rm res}$, are calculated by SPT\cite{Suda,Lekkerkerker} as follows: 
\begin{equation}
\beta p^{\rm res} v_{\rm bR}=\frac{\eta_{\rm lip}^{\rm res}q^{-2}}{\left(1-\eta_{\rm lip}^{\rm res} \right)^2}.
\label{Spres}
\end{equation}
$\alpha$ for the fluid and ordered phases are calculated by SPT\cite{Suda,Lekkerkerker} as follows:

\begin{eqnarray}
\alpha_{\rm f}&=&\left( 1-\eta_{\rm bR}^{\rm fluid}  \right){\rm exp}\left[- \frac{2\eta_{\rm bR}^{\rm fluid} q}{1- \eta_{\rm bR}^{\rm fluid} }  -\frac{\eta_{\rm bR}^{\rm fluid} q^2 }{\left(1-\eta_{\rm bR}^{\rm fluid} \right)^2} \right],\\
\alpha_{\rm ord}&=&\left( 1-\eta_{\rm bR}^{\rm ordered}  \right){\rm exp}\left[- \frac{2\eta_{\rm bR}^{\rm ordered} q}{1- \eta_{\rm bR}^{\rm ordered} }  -\frac{\eta_{\rm bR}^{\rm ordered} q^2 }{\left(1-\eta_{\rm bR}^{\rm ordered} \right)^2} \right].
\end{eqnarray}
The partial differentiations of $\alpha$ for the fluid and ordered phases are as follows: 
\begin{eqnarray}
\left( \frac{\partial \alpha_{\rm f}}{\partial \eta_{\rm bR}^{\rm fluid}}\right)=\frac{-(\eta_{\rm bR}^{\rm fluid})^2 +(-q^2+2q+2)\eta_{\rm bR}^{\rm fluid} -q^2-2q-1}{(1-\eta_{\rm bR}^{\rm fluid})^2}\nonumber \\
{\rm exp}\left[- \frac{2\eta_{\rm bR}^{\rm fluid} q}{1- \eta_{\rm bR}^{\rm fluid} }  -\frac{\eta_{\rm bR}^{\rm fluid} q^2 }{\left(1-\eta_{\rm bR}^{\rm fluid} \right)^2} \right],
\end{eqnarray}

\begin{eqnarray}
\left( \frac{\partial \alpha_{\rm ord}}{\partial \eta_{\rm bR}^{\rm ordered}}\right)=\frac{-(\eta_{\rm bR}^{\rm ordered})^2 +(-q^2+2q+2)\eta_{\rm bR}^{\rm ordered} -q^2-2q-1}{(1-\eta_{\rm bR}^{\rm ordered})^2} \nonumber \\
{\rm exp}\left[- \frac{2\eta_{\rm bR}^{\rm ordered} q}{1- \eta_{\rm bR}^{\rm ordered} }  -\frac{\eta_{\rm bR}^{\rm ordered} q^2 }{\left(1-\eta_{\rm bR}^{\rm ordered} \right)^2} \right].
\end{eqnarray}
To examine the convergence of the equations (\ref{Smu-mu}) and (\ref{Sp-p}) at the limit of $q=0$, $\alpha$ and $\left(\frac{\partial \alpha}{\partial \eta_{\rm bR}}\right)$ are expanded to polynomial functions of $q$. The Maclaurin expansions of $\alpha_{\rm f}$ and $\alpha_{\rm ord}$ are written as follows:
\begin{eqnarray}
\alpha_{\rm f}(q)=\alpha_{\rm f}(0)+\left(\frac{\partial \alpha_{\rm f}}{\partial q}\right)_{q=0}\cdot q+\frac{1}{2}\left(\frac{\partial^2 \alpha_{\rm f}}{\partial q^2}\right)_{q=0}\cdot q^2+O(q^3),\\
\alpha_{\rm ord}(q)=\alpha_{\rm ord}(0)+\left(\frac{\partial \alpha_{\rm ord}}{\partial q}\right)_{q=0}\cdot q+\frac{1}{2}\left(\frac{\partial^2 \alpha_{\rm ord}}{\partial q^2}\right)_{q=0}\cdot q^2+O(q^3).
\end{eqnarray}
$\alpha(0)$ for the fluid and ordered phases are as follows:
\begin{eqnarray}
\alpha_{\rm f}(0)&=& 1-\eta_{\rm bR}^{\rm fluid},\\
\alpha_{\rm ord}(0)&=& 1-\eta_{\rm bR}^{\rm ordered}.
\end{eqnarray}
$\left(\frac{\partial \alpha}{\partial q}\right)$ for the fluid and ordered phases are as follows:
\begin{eqnarray}
    \left(\frac{\partial \alpha_{\rm f}}{\partial q}\right)=\left(-2\eta_{\rm bR}^{\rm fluid}- \frac{2\eta_{\rm bR}^{\rm fluid}q}{(1-\eta_{\rm bR}^{\rm fluid})}  \right)\rm{exp}\left[\frac{-2\eta_{\rm bR}^{\rm fluid}\it q}{1-\eta_{\rm bR}^{\rm fluid}}- \frac{\eta_{\rm bR}^{\rm fluid}\it q^2}{(1-\eta_{\rm bR}^{\rm fluid})^2}  \right],\\
    \left(\frac{\partial \alpha_{\rm ord}}{\partial q}\right)=\left(-2\eta_{\rm bR}^{\rm ordered}- \frac{2\eta_{\rm bR}^{\rm ordered}q}{(1-\eta_{\rm bR}^{\rm ordered})}  \right)\rm{exp}\left[\frac{-2\eta_{\rm bR}^{\rm ordered}\it q}{1-\eta_{\rm bR}^{\rm ordered}}- \frac{\eta_{\rm bR}^{\rm ordered}\it q^2}{(1-\eta_{\rm bR}^{\rm ordered})^2}  \right].
\end{eqnarray}
$\left(\frac{\partial \alpha}{\partial q}\right)_{q=0}$ for the fluid and ordered phases are as follows:
\begin{eqnarray}
\left(\frac{\partial \alpha_{\rm f}}{\partial q}\right)_{q=0}&=&-2\eta_{\rm bR}^{\rm fluid},\\
\left(\frac{\partial \alpha_{\rm ord}}{\partial q}\right)_{q=0}&=&-2\eta_{\rm bR}^{\rm ordered}.
\end{eqnarray}
$\left(\frac{\partial^2 \alpha}{\partial q^2}\right)$ for the fluid and ordered phases are as follows:
\begin{eqnarray}
\left(\frac{\partial^2 \alpha_{\rm f}}{\partial q^2}\right)=(1-\eta_{\rm bR}^{\rm fluid})\left(\frac{-2\eta_{\rm bR}^{\rm fluid}}{(1-\eta_{\rm bR}^{\rm fluid})^2} +\left(\frac{-2\eta_{\rm bR}^{\rm fluid}}{1-\eta_{\rm bR}^{\rm fluid}}- \frac{2\eta_{\rm bR}^{\rm fluid}q}{(1-\eta_{\rm bR}^{\rm fluid})^2}  \right)^2\right)\nonumber\\
\rm{exp}\left[\frac{-2\eta_{\rm bR}^{\rm fluid}\it{q}}{1-\eta_{\rm bR}^{\rm fluid}}- \frac{\eta_{\rm bR}^{\rm fluid}\it{q}^2}{(1-\eta_{\rm bR}^{\rm fluid})^2}  \right],\\
\left(\frac{\partial^2 \alpha_{\rm ord}}{\partial q^2}\right)=(1-\eta_{\rm bR}^{\rm ordered})\left(\frac{-2\eta_{\rm bR}^{\rm ordered}}{(1-\eta_{\rm bR}^{\rm ordered})^2} +\left(\frac{-2\eta_{\rm bR}^{\rm ordered}}{1-\eta_{\rm bR}^{\rm ordered}}- \frac{2\eta_{\rm bR}^{\rm ordered}q}{(1-\eta_{\rm bR}^{\rm ordered})^2}  \right)^2\right)\nonumber\\
\rm{exp}\left[\frac{-2\eta_{\rm bR}^{\rm ordered}\it{q}}{1-\eta_{\rm bR}^{\rm ordered}}- \frac{\eta_{\rm bR}^{\rm ordered}\it{q}^2}{(1-\eta_{\rm bR}^{\rm ordered})^2}  \right].
\end{eqnarray}
$\left(\frac{\partial^2 \alpha}{\partial q^2}\right)_{q=0}$ for the fluid and ordered phases are as follows:
\begin{eqnarray}
\left( \frac{\partial^2 \alpha_{\rm f}}{\partial q^2}\right)_{q=0}&=&\frac{-2\eta_{\rm bR}^{\rm fluid}}{1-\eta_{\rm bR}^{\rm fluid}} +\frac{4\left(\eta_{\rm bR}^{\rm fluid}\right)^2}{1-\eta_{\rm bR}^{\rm fluid}},\\
\left( \frac{\partial^2 \alpha_{\rm ord}}{\partial q^2}\right)_{q=0}&=&\frac{-2\eta_{\rm bR}^{\rm ordered}}{1-\eta_{\rm bR}^{\rm ordered}} +\frac{4\left(\eta_{\rm bR}^{\rm ordered}\right)^2}{1-\eta_{\rm bR}^{\rm ordered}}.
\end{eqnarray}
The results of the Maclaurin expansions of $\alpha$ for the fluid and ordered phases are as follows:
\begin{eqnarray}
\alpha_{\rm f}(q)=(1-\eta_{\rm bR}^{\rm fluid})-2\eta_{\rm bR}^{\rm fluid}q
+\left(\frac{-\eta_{\rm bR}^{\rm fluid}+2\left(\eta_{\rm bR}^{\rm fluid}\right) ^2}{1-\eta_{\rm bR}^{\rm fluid}}\right)q^2+O(q^3),\\
\alpha_{\rm ord}(q)=(1-\eta_{\rm bR}^{\rm ordered})-2\eta_{\rm bR}^{\rm ordered}q
+\left(\frac{-\eta_{\rm bR}^{\rm ordered}+2\left(\eta_{\rm bR}^{\rm ordered}\right) ^2}{1-\eta_{\rm bR}^{\rm ordered}}\right)q^2
+O(q^3).
\end{eqnarray}

We define $\left(\frac{\partial \alpha}{\partial \eta_{\rm bR}}\right)$ as $G(q)$ for simple expression. The Maclaurin expansions of $G(q)$ for the fluid and ordered phases are as follows:
\begin{eqnarray}
G_{\rm f}(q)&=&G_{\rm f}(0)+\left( \frac{\partial G_{\rm f}}{\partial q}\right)_{q=0}\cdot q+\frac{1}{2}\left(\frac{\partial^2 G_{\rm f}}{\partial q^2}\right)_{q=0}\cdot q^2 +O(q^3),\\
G_{\rm ord}(q)&=&G_{\rm ord}(0)+\left( \frac{\partial G_{\rm ord}}{\partial q}\right)_{q=0}\cdot q+\frac{1}{2}\left(\frac{\partial^2 G_{\rm ord}}{\partial q^2}\right)_{q=0}\cdot q^2 +O(q^3).
\end{eqnarray}
We define $f(q)$ and $e(q)$ as follows:
\begin{eqnarray}
f(q)&\equiv& \frac{-\eta_{bR}^2 +(-q^2+2q+2)\eta_{bR}-q^2-2q-1}{(1-\eta_{bR})^2},\\
e(q)&\equiv& \rm exp\left[ -\frac{2\eta_{bR}\it{q}}{1-\eta_{bR}}-\frac{\eta_{bR}\it{q}^2}{(1-\eta_{bR})^2} \right],\\
G(q)&=&f(q)\cdot e(q).
\end{eqnarray}
$G(0)$ for the fluid and ordered phases are $-1$. $\left( \frac{\partial G}{\partial q}\right)$ for the fluid and ordered phases are as follows:
\begin{eqnarray}
    \left(\frac{\partial G_{\rm f}}{\partial q} \right)&=&\left(\frac{\partial f_{\rm f}}{\partial q} \right)\cdot e_{\rm f}(q)+f_{\rm f}(q)\cdot \left(\frac{\partial e_{\rm f}}{\partial q} \right),\\
    \left(\frac{\partial G_{\rm ord}}{\partial q} \right)&=&\left(\frac{\partial f_{\rm ord}}{\partial q} \right)\cdot e_{\rm ord}(q)+f_{\rm ord}(q)\cdot \left(\frac{\partial e_{\rm ord}}{\partial q} \right).
\end{eqnarray}
$\left(\frac{\partial f}{\partial q} \right)$ for the fluid and ordered phases are as follows:
\begin{eqnarray}
\left(\frac{\partial f_{\rm f}}{\partial q} \right)&=&\frac{(-2q+2)\eta_{\rm bR}^{\rm fluid}-2q-2}{(1-\eta_{\rm bR}^{\rm fluid})^2},\\
\left(\frac{\partial f_{\rm ord}}{\partial q} \right)&=&\frac{(-2q+2)\eta_{\rm bR}^{\rm ordered}-2q-2}{(1-\eta_{\rm bR}^{\rm ordered})^2}.
\end{eqnarray}
$\left(\frac{\partial e}{\partial q} \right)$ for the fluid and ordered phases are as follows:
\begin{eqnarray}
\left(\frac{\partial e_{\rm f}}{\partial q} \right)=\left( -\frac{2\eta_{\rm bR}^{\rm fluid}}{1-\eta_{\rm bR}^{\rm fluid}}\ -\frac{2\eta_{\rm bR}^{\rm fluid}q}{(1-\eta_{\rm bR}^{\rm fluid})^2} \right)\rm{exp}\left[  -\frac{2\eta_{\rm bR}^{\rm fluid}\it{q}}{1-\eta_{\rm bR}^{\rm fluid}}\ -\frac{\eta_{\rm bR}^{\rm fluid}\it{q}^2}{(1-\eta_{\rm bR}^{\rm fluid})^2} \right],\\
\left(\frac{\partial e_{\rm ord}}{\partial q} \right)=\left( -\frac{2\eta_{\rm bR}^{\rm ordered}}{1-\eta_{\rm bR}^{\rm ordered}}\ -\frac{2\eta_{\rm bR}^{\rm ordered}q}{(1-\eta_{\rm bR}^{\rm ordered})^2} \right)\rm{exp}\left[  -\frac{2\eta_{\rm bR}^{\rm ordered}\it{q}}{1-\eta_{\rm bR}^{\rm ordered}}\ -\frac{\eta_{\rm bR}^{\rm ordered}\it{q}^2}{(1-\eta_{\rm bR}^{\rm ordered})^2} \right].
\end{eqnarray}
$\left( \frac{\partial G}{\partial q}\right)_{q=0}$ for the fluid and ordered phases are as follows:
\begin{eqnarray}
\left( \frac{\partial G_{\rm f}}{\partial q}\right)_{q=0} &=& -\frac{2}{1-\eta_{\rm bR}^{\rm fluid}}\cdot 1-1\cdot \left( -\frac{2\eta_{\rm bR}^{\rm fluid}}{1-\eta_{\rm bR}^{\rm fluid}} \right)=-2,\\
\left( \frac{\partial G_{\rm ord}}{\partial q}\right)_{q=0} &=& -\frac{2}{1-\eta_{\rm bR}^{\rm ordered}}\cdot 1-1\cdot \left( -\frac{2\eta_{\rm bR}^{\rm ordered}}{1-\eta_{\rm bR}^{\rm ordered}} \right)=-2.
\end{eqnarray}
$\left( \frac{\partial^2 G}{\partial q^2}\right)$ for the fluid and ordered phases are as follows:
\begin{eqnarray}
  \left( \frac{\partial^2 G_{\rm f}}{\partial q^2}\right)=\left( \frac{\partial^2 f_{\rm f}}{\partial q^2}\right)\cdot e_{\rm f}(q)+2\left( \frac{\partial f_{\rm f}}{\partial q}\right)\cdot \left( \frac{\partial e_{\rm f}}{\partial q}\right)+f_{\rm f}(q)\cdot \left( \frac{\partial^2 e_{\rm f}}{\partial q^2}\right),\\
  \left( \frac{\partial^2 G_{\rm ord}}{\partial q^2}\right)=\left( \frac{\partial^2 f_{\rm ord}}{\partial q^2}\right)\cdot e_{\rm ord}(q)+2\left( \frac{\partial f_{\rm ord}}{\partial q}\right)\cdot \left( \frac{\partial e_{\rm ord}}{\partial q}\right)+f_{\rm ord}(q)\cdot \left( \frac{\partial^2 e_{\rm ord}}{\partial q^2}\right).
\end{eqnarray}
$\left(\frac{\partial^2 f}{\partial q^2} \right)$ for the fluid and ordered phases are as follows:
\begin{eqnarray}
    \left(\frac{\partial^2 f_{\rm f}}{\partial q^2} \right)&=&\frac{-2\eta_{\rm bR}^{\rm fluid}-2}{(1-\eta_{\rm bR}^{\rm fluid})^2},\\
    \left(\frac{\partial^2 f_{\rm ord}}{\partial q^2} \right)&=&\frac{-2\eta_{\rm bR}^{\rm ordered}-2}{(1-\eta_{\rm bR}^{\rm ordered})^2}.
\end{eqnarray}
$\left(\frac{\partial^2 e}{\partial q^2} \right)$ for the fluid and ordered phases are as follows:
\begin{eqnarray}
\left(\frac{\partial^2 e_{\rm f}}{\partial q^2} \right)&=&\left(\frac{-2\eta_{\rm bR}^{\rm fluid}}{(1-\eta_{\rm bR}^{\rm fluid})^2}+\left( -\frac{2\eta_{\rm bR}^{\rm fluid}}{1-\eta_{\rm bR}^{\rm fluid}}\ -\frac{2\eta_{\rm bR}^{\rm fluid}q}{(1-\eta_{\rm bR}^{\rm fluid})^2} \right)^2 \right) \nonumber\\
&\rm{exp}&\left[  -\frac{2\eta_{\rm bR}^{\rm fluid}\it{q}}{1-\eta_{\rm bR}^{\rm fluid}}\ -\frac{\eta_{\rm bR}^{\rm fluid}\it{q}^2}{(1-\eta_{\rm bR}^{\rm fluid})^2} \right],\\
\left(\frac{\partial^2 e_{\rm ord}}{\partial q^2} \right)&=&\left(\frac{-2\eta_{\rm bR}^{\rm ordered}}{(1-\eta_{\rm bR}^{\rm ordered})^2}+\left( -\frac{2\eta_{\rm bR}^{\rm ordered}}{1-\eta_{\rm bR}^{\rm ordered}}\ -\frac{2\eta_{\rm bR}^{\rm ordered}q}{(1-\eta_{\rm bR}^{\rm ordered})^2} \right)^2 \right) \nonumber\\
&\rm{exp}&\left[  -\frac{2\eta_{\rm bR}^{\rm ordered}\it{q}}{1-\eta_{\rm bR}^{\rm ordered}}\ -\frac{\eta_{\rm bR}^{\rm ordered}\it{q}^2}{(1-\eta_{\rm bR}^{\rm ordered})^2} \right].
\end{eqnarray}
$\left( \frac{\partial^2 G}{\partial q^2}\right)_{q=0}$ for the fluid and ordered phases are as follows:
\begin{eqnarray}
    \left( \frac{\partial^2 G_{\rm f}}{\partial q^2}\right)_{q=0}&=&\frac{-2\eta_{\rm bR}^{\rm fluid}-2}{(1-\eta_{\rm bR}^{\rm fluid})^2}+2\left(\frac{-2}{(1-\eta_{\rm bR}^{\rm fluid})}\cdot \frac{-2\eta_{\rm bR}^{\rm fluid}}{(1-\eta_{\rm bR}^{\rm fluid})}\right)+-1\cdot \frac{-2\eta_{\rm bR}^{\rm fluid}+4\left(\eta_{\rm bR}^{\rm fluid}\right)^2}{(1-\eta_{\rm bR}^{\rm fluid})^2}\nonumber\\
    &=&\frac{-2+8\eta_{\rm bR}^{\rm fluid}-4\left(\eta_{\rm bR}^{\rm fluid}\right)^2}{(1-\eta_{\rm bR}^{\rm fluid})^2},\\
    \left( \frac{\partial^2 G_{\rm ord}}{\partial q^2}\right)_{q=0}&=&\frac{-2\eta_{\rm bR}^{\rm ordered}-2}{(1-\eta_{\rm bR}^{\rm ordered})^2}+2\left(\frac{-2}{(1-\eta_{\rm bR}^{\rm ordered})}\cdot \frac{-2\eta_{\rm bR}^{\rm ordered}}{(1-\eta_{\rm bR}^{\rm ordered})}\right)+\nonumber\\
    &-1&\cdot \frac{-2\eta_{\rm bR}^{\rm ordered}+4\left(\eta_{\rm bR}^{\rm ordered}\right)^2}{(1-\eta_{\rm bR}^{\rm ordered})^2}\nonumber\\
    &=&\frac{-2+8\eta_{\rm bR}^{\rm ordered}-4\left(\eta_{\rm bR}^{\rm ordered}\right)^2}{(1-\eta_{\rm bR}^{\rm ordered})^2}.
\end{eqnarray}
The results of the Maclaurin expansions of $G(q)$ for the fluid and ordered phases are as follows
\begin{eqnarray}
G_{\rm f}(q)&=&-1-2q+\frac{-1+4\eta_{\rm bR}^{\rm fluid}-2\left(\eta_{\rm bR}^{\rm fluid}\right)^2}{(1-\eta_{\rm bR}^{\rm fluid})^2} q^2 +O(q^3),\label{SGf}\\
G_{\rm ord}(q)&=&-1-2q+\frac{-1+4\eta_{\rm bR}^{\rm ordered}-2\left(\eta_{\rm bR}^{\rm ordered}\right)^2}{(1-\eta_{\rm bR}^{\rm ordered})^2} q^2 +O(q^3).\label{SGord}
\end{eqnarray}
The $\alpha - \eta_{\rm bR}\left( \frac{\partial \alpha}{\partial \eta_{\rm bR}}\right)$ for the fluid and ordered phases are as follows:
\begin{eqnarray}
\alpha_{\rm f}-\eta_{\rm bR}^{\rm fluid}\left( \frac{\partial \alpha_{\rm f}}{\partial \eta_{\rm bR}^{\rm fluid}}\right)&=&1-\frac{\left(\eta_{\rm bR}^{\rm fluid}\right)^2}{(1-\eta_{\rm bR}^{\rm fluid})^2}q^2+O(q^3),\label{Saf-Gf}\\
\alpha_{\rm ord}-\eta_{\rm bR}^{\rm ordered}\left( \frac{\partial \alpha_{\rm ord}}{\partial \eta_{\rm bR}^{\rm ordered}}\right)&=&1-\frac{\left(\eta_{\rm bR}^{\rm ordered}\right)^2}{(1-\eta_{\rm bR}^{\rm ordered})^2}q^2+O(q^3).\label{Saord-Gord}
\end{eqnarray}
Substituting the eqs. (\ref{Spres}), (\ref{SGf}), and (\ref{SGord}) to (\ref{Smu-mu}), the eq (\ref{Smu-mu}) is calculated as
\begin{equation}
\mu^0_{\rm f}-\mu_{\rm ord}^0+\frac{\eta_{\rm lip}^{\rm res}}{(1-\eta_{\rm lip}^{\rm res})^2}\cdot \left( \frac{1-4\eta_{\rm bR}^{\rm fluid}+2\left(\eta_{\rm bR}^{\rm fluid}\right)^2}{(1-\eta_{\rm bR}^{\rm fluid})^2}-\frac{1-4\eta_{\rm bR}^{\rm ordered}+2\left(\eta_{\rm bR}^{\rm ordered}\right)^2}{(1-\eta_{\rm bR}^{\rm ordered})^2} \right)+O(q)=0.\label{SMmu-mu}
\end{equation}
Substituting the eqs. (\ref{Spres}), (\ref{Saf-Gf}), and (\ref{Saord-Gord}) to (\ref{Sp-p}), the eq (\ref{Sp-p}) is calculated as
\begin{eqnarray}
    &(p^0_{\rm f}-p^0_{\rm ord})&v_{\rm bR}+\frac{\eta_{\rm lip}^{\rm res}}{(1-\eta_{\rm lip}^{\rm res})^2}\cdot \left( \frac{-(\eta_{\rm bR}^{\rm fluid})^2}{ (1-\eta_{\rm bR}^{\rm fluid})^2}+\frac{(\eta_{\rm bR}^{\rm ordered})^2}{ (1-\eta_{\rm bR}^{\rm ordered})^2}  \right)\nonumber\\
    &+O(q)=0&.\label{SMp-p}
\end{eqnarray}
The eqs. (\ref{SMmu-mu}) and (\ref{SMp-p}) show that the solutions of eqs. (\ref{Smu-mu}) and (\ref{Sp-p}) are independent of $q$ at the limit of $q=0$. The phase diagram was obtained using the equations (\ref{SMmu-mu}) and (\ref{SMp-p}). The coexistence region of this phase diagram almost overlap that of phase diagrams at very small $q$, $q=0.001$ and $q=0.0001$ (data is not shown). 





\end{document}